\documentclass[a4paper,fleqn]{cas-sc}

\usepackage[numbers]{natbib}
\usepackage{graphicx}%
\usepackage{multirow}%
\usepackage{amsmath,amssymb,amsfonts}%
\usepackage{mathtools}
\usepackage{amsthm}%
\usepackage{mathrsfs}%
\usepackage[title]{appendix}%
\usepackage{xcolor}%
\usepackage{textcomp}%
\usepackage{manyfoot}%
\usepackage{booktabs}%
\usepackage{algorithm}%
\usepackage{algorithmicx}%
\usepackage{algpseudocode}%
\usepackage{listings}%
\usepackage{ulem}
\usepackage{cancel}

\definecolor{magenta}{rgb}{0.8, 0.1, 0.6}

\def\v#1{{\boldsymbol{#1}}} 
\def\m#1{{\mathcal{#1}}} 

\def\tsc#1{\csdef{#1}{\textsc{\lowercase{#1}}\xspace}}
\tsc{WGM}
\tsc{QE}
\tsc{EP}
\tsc{PMS}
\tsc{BEC}
\tsc{DE}

\begin{document}
\let\WriteBookmarks\relax
\def\floatpagepagefraction{1}
\def\textpagefraction{.001}
\shorttitle{Phase changes of the flow rate in the VA caused by dTEVAR}
\shortauthors{N. Takeishi \& N. Yokoyama et~al.}

\title[mode = title]{Phase changes of the flow rate in the vertebral artery caused by debranching thoracic endovascular aortic repair: effects of flow path and local vessel stiffness on vertebral arterial pulsation}

\author[1]{Naoki Takeishi}[
   type=editor,
   auid=000, bioid=1,
   orcid=0000-0002-9568-8711
   ]
\cormark[1]
\ead{takeishi.naoki.008@m.kyushu-u.ac.jp}

\author[2]{Li Jialong}[
   ]

\author[3]{Naoto Yokoyama}[
   orcid=0000-0003-1460-1002
   ]
\cormark[1]
\ead{n.yokoyama@mail.dendai.ac.jp}

\author[4]{Takasumi Goto}
\author[5]{Hisashi Tanaka}
\author[4]{Shigeru Miyagawa}
\author[2]{Shigeo Wada}


\address[1]{Department of Mechanical Engineering, Faculty of Engineering, Kyushu University, 744 Motooka, Nishi-ku, Fukuoka 819-0395,
Japan}
\address[2]{Graduate School of Engineering Science, Osaka University, 1-3 Machikaneyama, Toyonaka, 560-8531, Osaka, Japan}
\address[3]{Department of Mechanical Engineering, Tokyo Denki University, 5 Senju-Asahi, Adachi, 120-8551, Tokyo, Japan}
\address[4]{Department of Cardiovascular Surgery, Osaka University Graduate School of Medicine, Suita, Osaka, 565-0871, Japan}
\address[5]{Graduate School of Medicine, Division of Health Science, Osaka University, 2-2 Yamadaoka, Suita, Osaka 565-0871, Japan}

\cortext[cor1]{Corresponding author}


\begin{abstract}
Despite numerous studies on cerebral arterial blood flow,
there has not yet been a comprehensive description of hemodynamics in patients undergoing debranching thoracic endovascular aortic repair (dTEVAR),
a promising surgical option for aortic arch aneurysms.
A phase delay of the flow rate in the left vertebral artery (LVA) in patients after dTEVAR compared to those before was experimentally observed, 
while the phase in the right vertebral artery (RVA) remained almost the same before and after surgery.
Since this surgical intervention included stent graft implantation and extra-anatomical bypass,
it was expected that the intracranial hemodynamic changes due to dTEVAR were coupled with fluid flow and pulse waves in cerebral arteries.
To clarify this issue,
A one-dimensional model ($1$D) was used to numerically investigate the relative contribution (i.e., local vessel stiffness and flow path changes) of the VA flow rate to the phase difference.
The numerical results demonstrated a phase delay of flow rate in the LVA but not the RVA in postoperative patients undergoing dTEVAR relative to preoperative patients.
The results further showed that the primary factor affecting the phase delay of the flow rate in the LVA after surgery compared to that before was the bypass, i.e., alteration of flow path,
rather than stent grafting, i.e., the change in local vessel stiffness.
The numerical results provide insights into hemodynamics in postoperative patients undergoing dTEVAR,
as well as knowledge about therapeutic decisions.
\end{abstract}




\maketitle

\section{Introduction}\label{sec1}

Aortic aneurysm, which is morphologically defined as the focal dilation and structural degradation of the aorta,
is an asymptomatic disease,
and its rupture is a significant cause of death worldwide.
For instance, in the United States of America, aortic aneurysms and dissections cause over $10,000$ deaths each year~\cite{Go2013, Milewicz2019}.
To date, thoracic endovascular aortic repair (TEVAR) is the only effective treatment for aortic aneurysm.
TEVAR has been a particularly attractive surgical intervention because it is less invasive than conventional open surgical repair~\cite{Goto2019, Narita2016, Shijo2016, Shimamura2008}.

The aorta is the largest conduit artery in the body, and due to its extraordinary ability to expand and contract, it serves as a reservoir that transforms the highly pressured and pulsatile heart output into a flow with moderate fluctuations~\cite{Ku1997}.
Deterioration of this buffering effect due to stiffening of the arterial wall is the main cause of hypertension~\cite{Ku1997, Liu1989}.
These dynamics of the aorta are also important for propagation of the pulse wave,
which has been recognized as an early indicator of the health status of the cardiovascular system~\cite{Hametner2021, Laurent2006, Townsend2015}.
Thus, understanding the hemodynamic differences between pre- and postoperative TEVAR patients based on pulse-wave dynamics is of paramount importance, not only for surgical decision-making to achieve optimal clinical outcomes, but also for evaluating postoperative hemodynamics.

In TEVAR, the aortic arch is mapped according to the segmentation of the vertebral column or landing zone~\cite{Ishimaru2004},
where zone $0$ ranges from the ascending aorta (AA) to the origin of the innominate artery (IA),
zone $1$ ranges from the IA to the origin of the left common carotid artery (LCCA),
zone $2$ ranges from the LCCA to the left subclavian artery (LSA),
and zones 3 and 4 follow in the longitudinal direction.
TEVAR for partial arch debranching,
known as debranching TEVAR (dTEVAR),
includes both stent graft implantation and extra-anatomical bypass,
where the stent graft is placed from zone $1$ or $2$ with a length of several millimeters depending on the patient.
This operation is most often performed in high-risk patients with thoracic aortic aneurysms (TAA)~\cite{Narita2016, Shijo2016, Shimamura2008}.
Currently, expanded polytetrafluoroethylene (ePTFE) and polyethylene therephthalate (PET) are the two most popular graft materials in abdominal aortic aneurysm repair due to their remarkable biocompatibility and durability.
Additional supra-arch vessel reconstruction in dTEVAR is known to prevent cerebral infarction~\cite{Hajibandeh2016, Shigemura2000, Ullery2012, Yoshitake2016} and ischemia in the cerebral circulation~\cite{Cooper2009, Feezor2009, Patterson2014}.

A previous experimental study has shown that total intracranial blood flow was preserved after dTEVAR,
with significant decrease of flow in the LVA and significant increase of flow in the RVA~\cite{Goto2019}.
The hemodynamic mechanism of such changes in posterior cerebral circulation after dTEVAR remains uncertain yet due the technical difficulty of measurements.
Thus, our primary concern here is to evaluate bilateral vertebral artery blood flows under pulsation between the patients before and after dTVEAR with axillo-axillary artery (AxA-AxA) bypass,
and to clarify the hemodynamic mechanism associated with postoperative significant changes of posterior cerebral circulation. 
Since dTEVAR includes extra-anatomical bypass,
it was expected that the intracranial hemodynamic changes due to dTEVAR would be coupled with fluid flow and pulse waves in the VA.
Therefore, the objective of this study was to numerically clarify whether there exist phase differences of the flow rate in the VA,
as a representative intracranial vessel,
between patients before and after dTEVAR,
especially in cases with one bypass from the right axillary artery (RAxA) to the left axillary artery (LAxA) and embolization at the branch point between a stent-grafted aortic arch and the LSA (see Figure~\ref{fig:schematics}).
If phase differences of the flow rate exist,
we clarify how surgery contributes to it.
Hereafter, we refer to this operation as single-debranching TEVAR ($1$dTEVAR).
A one-dimensional ($1$D) model analysis was used to numerically investigate blood flow rates in the L/RVA in pre- and postoperative patients after $1$dTEVAR.

In a human aorta with $\sim10$ mm radius and $\sim0.3$ mm thickness,
the wave speed, quantified by the traditional foot-to-foot wave velocity method, e.g., by~Laurent et al.~\cite{Laurent2006}, is over $5$ m/s~\cite{Latham1985}.
Such conventional experimental methodology, however, is still difficult to quantify the effect of pulse waves in the VA on the intracranial blood supply.
Furthermore, assuming that one pulse has a duration of $1$ second,
the ratio between the wave length and radius is $500$, i.e., long wave.
Numerical analysis for blood flow with long-wave pulses still involves a heavy computational load even with a two-dimensional model,
although several attempts for this issue were reported, e.g., in~\cite{Yokoyama2021}.
$1$D model analysis is one of the most effective and practical non-invasive solutions to investigate hemodynamics with long-wave pulses~\cite{Formaggia2003, Taylor2004, vandeVosse2011}.
Thus, for practical purposes, 1D modeling has been widely used in circulatory systems,
such as the systemic arteries~\cite{Alastruey2011, Olufsen1999},
coronary circulation~\cite{Duanmu2019, Huo2007, Mynard2014},
the circle of Willis (CoW)~\cite{Alastruey2007},
and large vasculature, including venous systems coupled with contributions of the microcirculation modeled by lumped parameters~\cite{Duanmu2019, Huo2007, Mynard2014, Alastruey2008, Liang2009, Mynard2012}.
Recently, $1$D modeling was applied to a problem involving both blood flow rate and blood oxygenation~\cite{Feiger2020}.
Padmos et al.~\cite{Padmos2020} used patient-specific medical imaging data on the anatomy of the CoW to perform a $1$D model analysis of intracranial blood flow assuming a steady state,
coupled with three-dimensional tissue perfusion.
In this study, 
the experimental evidence on phase differences in the flow rate in the L/RVA between patients before and after $1$dTEVAR was reported.
Next, using 1D numerical model, the phase difference of the flow rate in the L/RVA in pre- and postoperative patients was evaluated.
Simulations were also performed for different local vessel stiffnesses and lengths of vessels that had stiffened due to stent grafting.

\section{Materials and Methods}

\subsection{Subjects and measurements}
\begin{figure*}[htbp!]
  \begin{center}
  \includegraphics[width=8cm]{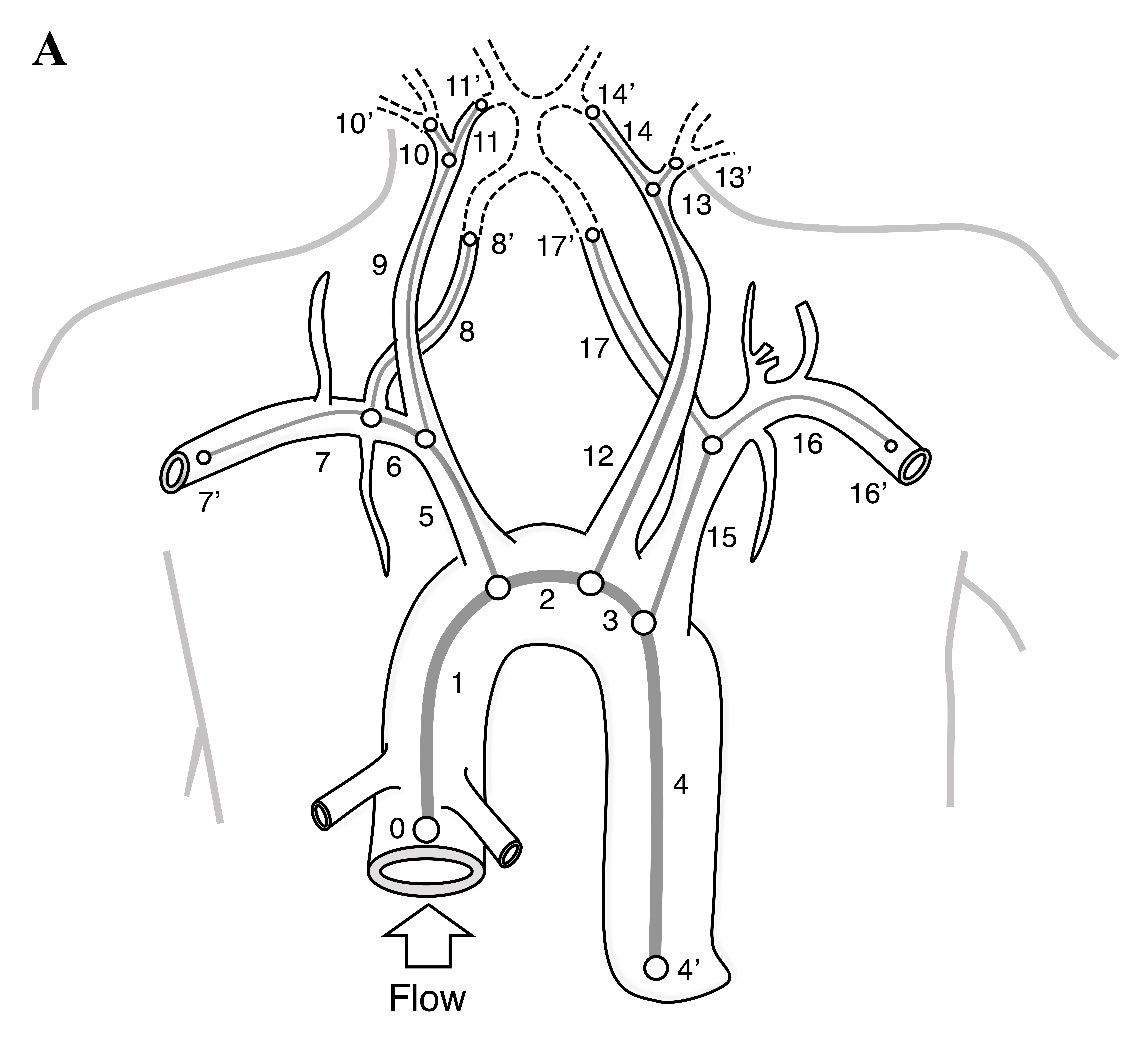}
  \includegraphics[width=8cm]{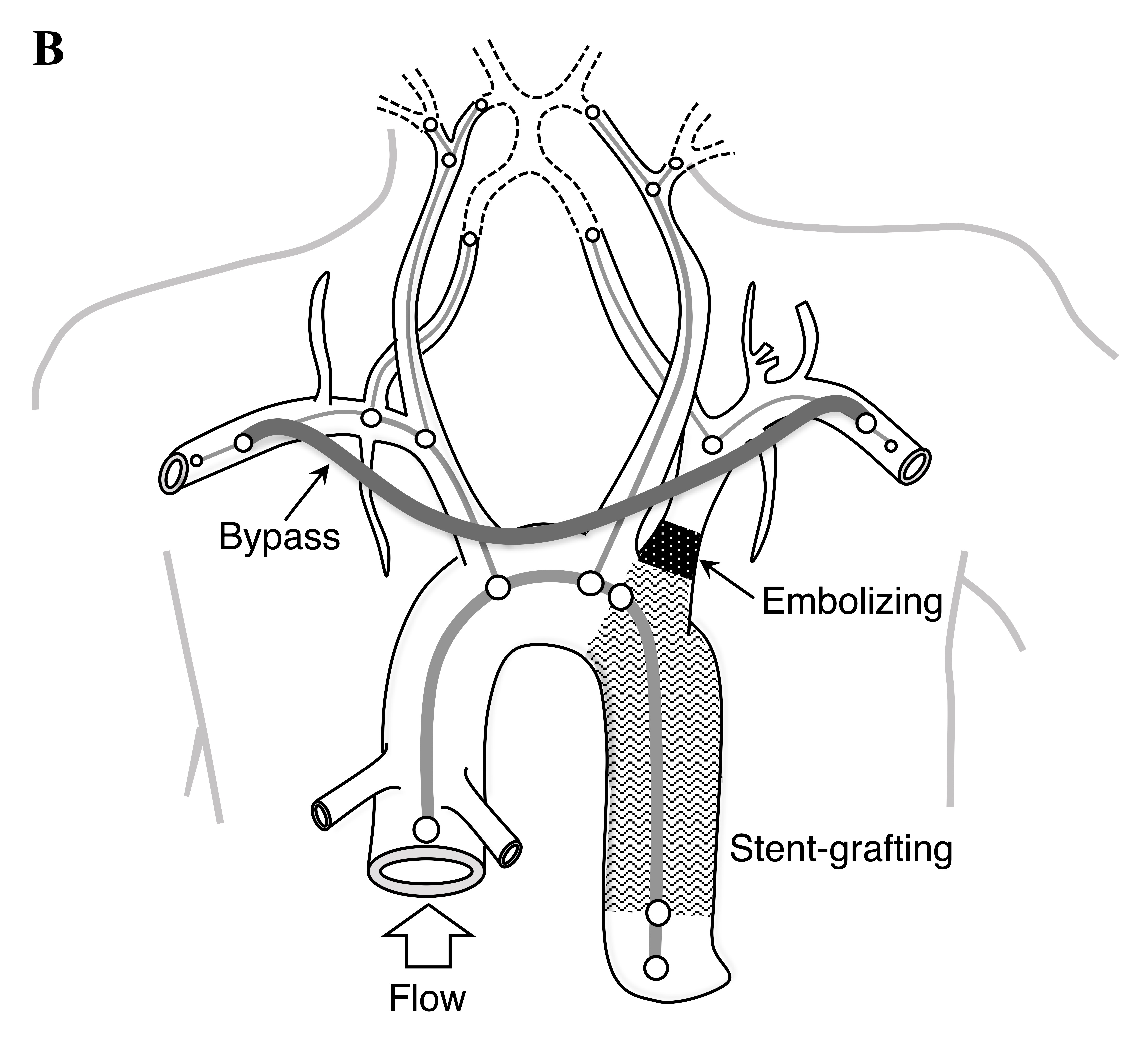}
  \end{center}
  \caption{Schematics of the cervical vasculature in \textbf{(A)} a normal subject (before surgery) and \textbf{(B)} a patients after $1$dTEVAR with AxA-AxA bypass.
  Blood flow starts from the inlet (LV, ID $= 0$) and continues through the AA (ID $= 1$) into all bilateral vessels, including the ICA (ID $= 11$ and $14$),
  VA (ID $= 8$ and $17$),
  and axillary arteries (ID $= 7$ and $16$).
  Superscript `` ' '' denotes the end terminal of each artery.}
  \label{fig:schematics}
\end{figure*}

This retrospective study was conducted in accordance with the guidelines of the Declaration of Helsinki.
All experimental protocols were approved by the institutional review board of Osaka University.
All subjects provided oral and written informed consent to participate in this study.
A prospective analysis of blood flow in the left and right internal carotid arteries (L/RICA) and the L/RVA in 9 patients before and after dTEVAR was performed between January $2015$ and January $2020$ at Osaka University Hospital,
where RAxA-LAxA bypass was performed (see also Figure~\ref{fig:schematics}B).
All patients were male,
with a mean age of $73.7\pm14.4$ years (standard deviation, SD).
The characteristics of the enrolled patients are presented in Table~1. 
Note that a magnetic resonance imaging (MRI) confirmed no stenosis in the CoW in preoperative patients.
A ringed $8$-mm ePTFE graft (FUSION, MAQUET Getinge Group, Japan) was used in all the debranching procedures (see also~\cite{Goto2019}).
Two-dimensional ($2$D) cine phase-contrast data was acquired with $3$T MRI (MR$750$-$3$T, GE Healthcare, Waukesha, WI, USA).
Validations of the resolution of $3$T MRI have been performed in {\it in vivo} studies, for instance, by~Calderon-Arnulphi et al.~\cite{Calderon-Arnulphi2011} and Zhao et al.~\cite{Zhao2007}.
In all patients, MRI was performed within $6$ months prior to the 1dTEVAR procedure and within $1$ month afterward~\cite{Goto2019}.
\begin{table*}[h!]
  \begin{center}
    \caption{Patient characteristics.}
   \label{tab:character}
    \begin{tabular}{l c} \hline
        Number of patients, $n$ & $9$ \\
        Mean age (SD) & $73.7 (14.4)$ \\
        Gender (men$:$women) & $9:0$ \\
        Current smoker, $n$ ($\%$) & $5/9 (55.6\%)$ \\        
        Comorbidities, $n$ ($\%$) & \\
        \hspace{2em} Cardiovascular disease & $6/9 (66.7\%)$ \\
        \hspace{2em} COPD & $2/9 (22.2\%)$ \\ 
        \hspace{2em} Chronic kidney disease & $3/9 (33.3\%)$ \\
        \hspace{2em} Haemodialysis & $0/9 (0\%)$ \\        
        \hspace{2em} Diabetes mellitus & $2/9 (22.2\%)$ \\        
        \hspace{2em} Hypertension & $9/9 (100\%)$ \\
        \hspace{2em} Dyslipidaemia & $5/9 (55.6\%)$ \\        
        \hspace{2em} Cerebrovascular disease (Old cerebral infarction) & $2/9 (22.2\%)$ \\
      \hline
    \end{tabular}
  \end{center}
  COPD: Chronic obstructive pulmonary disease
\end{table*}

The vascular tree models of both pre- and postoperative patients were reconstructed based on geometrical parameters of vessel diameter and length between branch points, as shown in Figure~\ref{fig:schematics}.
Arterial vascular geometries, including reference radius ($r_0$) and length ($L$) in a representative patient, were measured both pre- and postoperatively at Osaka University Hospital,
and the data are summarized in Table~\ref{tab:vessels}.
The diameters of the end terminals,
which are denoted by the vessel ID $= \chi^\prime$ in Figure~\ref{fig:schematics}\textbf{(A)},
were determined so that the time-average flow rate became similar to that obtained with experimental measurements by Bogren \& Buonocore~\cite{Bogren1994} or Goto et al.~\cite{Goto2019} (see Figure~\ref{fig:comparison}).
The lengths of the end terminals were uniformly set to $50$ mm.
For simplicity, branch angles were uniformly set to $30$ deg.
The effect of branch angle is mentioned in Appendix~\S{A.1} (see Figure~\ref{fig:results_angle}).
To represent arterial trees in patients after $1$dTEVAR,
the Young's modulus was set to be almost $100$ times larger in the stent-grafted area (vessel ID $= 4$; see Table~\ref{tab:vessels}) than in preoperative patients,
and also created a bypass from the RAxA to the LAxA (vessel ID $= 7$ and $16$) with embolization at the branch point from the stent-grafted thoracic aorta to the LSA (see Figure~\ref{fig:schematics}\textbf{(B)}).
In this study,
the bypass was considered to have a radius of $4.5$ mm and a length of $262$ mm,
and the stent-grafted artery was considered to have a radius of $12.5$ mm and a length of $40$ mm.
The diameters and lengths of end terminals were also the same as those in the preoperative vascular tree model.
\begin{table*}[h!]
  \begin{center}
    \caption{Geometrical parameters ($r_0$ and $L$) in a representative patient.}
   \label{tab:vessels}
    \begin{tabular}{c l c c} \hline
          ID  &  Artery & Radius $r_0$ [mm] & Length $L$ [mm] \\ \hline
        $0$ & Left Ventricle (LV; inlet boundary) & $12.5$ & - \\ 
        $1$ & Ascending Aorta (AA) & $12.5$ & $50$ \\ 
        $2$ & Aortic Arch I & $12.5$ & $20$ \\ 
        $3$ & Aortic Arch II & $12.5$ & $10$ \\
        $4$ &Thoracic Aorta & $12.5$ & $30$ \\ 
        $5$ & Brachiocephalic & $7$ & $33$ \\ 
        $6$ & Right Subclavian (RSA)& $4$ & $54$ \\ 
        $7$ & Right Axillary & $4$ & $139$ \\ 
        $8$ & Right Vertebral (RVA) & $1.5$ & $200$ \\ 
        $9$ & Right Carotid  & $2.5$ & $100$ \\ 
      $10$ & Right External Carotid (RECA) & $2$ & $50$ \\ 
      $11$ & Right Internal Carotid (RICA) & $2$ & $100$ \\ 
      $12$ & Left Carotid  & $2.5$ & $130$ \\ 
      $13$ & Left External Carotid (LECA) & $2$ & $50$ \\ 
      $14$ & Left Internal Carotid (LICA) & $2$ & $100$ \\ 
      $15$ & Left Subclavian (LSA)& $4$ & $60$ \\ 
      $16$ & Left Axillary & $4$ & $150$ \\ 
      $17$ & Left Vertebral (LVA) & $1.5$ & $200$ \\ 
      \hline
      $-$ & Bypass & $4.5$ & $262$ \\
      $-$ & Stent-Grafted Region & $12.5$ & $40$ (--$240$) \\
      \hline
    \end{tabular}
  \end{center}
\end{table*}

\subsection{Mathematical model and simulation}
The $1$D governing equations describe the conservation of mass and momentum:
\begin{subequations}
\begin{align}
&\frac{\partial A}{\partial t} + \frac{\partial Q}{\partial x} = 0, \\
&\frac{\partial Q}{\partial t} + \frac{\partial}{\partial x} \left( \alpha \frac{Q^2}{A} \right) + \frac{A}{\rho} \frac{\partial p}{\partial x} + K_R \frac{Q}{A} = 0,
\end{align}
\label{eq:govern}
\end{subequations}
where $x$ is the axial direction,
$A = A (x, t)$ is the area of a cross-section at $x$ and at time $t$,
$Q (= UA)$ is the mean volumetric flow rate across a section,
$U = U (x, t)$ is the velocity of the fluid averaged across the section,
$p = p (x, t)$ is the pressure,
$\rho$ ($= 1.05 \times 10^3$ kg/m$^3$) is the blood density,
$\alpha$ is the coefficient of the velocity profile,
and $K_R$ ($= 22 \pi \m{V} = 22 \pi \mu/\rho$) is the drag coefficient for the blood viscosity $\mu$ ($= 4.5 \times 10^{-3}$ Pa$\cdot$s).
A flat velocity profile was assumed, and set to be $\alpha = 1$~\cite{Sherwin2003}.
Assuming static equilibrium in the radial direction of a cylindrical tube or thin elastic shell,
one can derive a pressure relationship of the form~\cite{Olufsen1999}
\begin{equation}
  p - p_\mathrm{ext} = \beta \left( \sqrt{A} - \sqrt{A_0} \right),
\end{equation}
and
\begin{align}
  \beta &= \frac{\sqrt{\pi} h_0 E(x)}{\left( 1 - \nu^2 \right) A_0}, \label{eq:beta}\\
  E(x) &= \frac{r_0}{h_0} \left[ k_1 \exp{\left( k_2 r_0 \right)} + k_3 \right], \label{eq:E}
\end{align}
where $h_0 = h_0 (x)$, $r_0 = r_0 (x)$, and $A_0 = \pi r_0^2$ are the vessel thickness, vessel radius, and sectional area, respectively, at the equilibrium state $(p, Q) = (p_\mathrm{ext}, 0)$,
$E(x)$ is the Young's modulus,
$p_\mathrm{ext} (= 0)$ is the external pressure, assumed as a constant,
$\nu$ is the Poisson ratio, which is set to be $\nu = 0.5$ for practical incompressibility, and
$k_i$ ($i = 1$--$3$) are the coefficients set to $k_1 = 2 \times 10^6$ kg/(s$^2$$\cdot$m), $k_2 = -2.253 \times 10^3$ 1/m, and $k_3 = 8.65 \times 10^4$ kg/(s$^2$$\cdot$m)~\cite{Olufsen1999, Taylor2004}.
These parameter values were also used in previous 1D blood flow analyses, e.g., by Alastruey et al.~\cite{Alastruey2011} and Smith~\cite{Smith2002}.
Since the effect of $h_0$ does not appear in $\beta$ owing to equations~\eqref{eq:beta} and \eqref{eq:E},
and since the vessel stiffness is simply characterized by $\beta$, which is determined mainly by the reference radius $r_0$ and parameters $k_i$,
the value of $h_0$ in each artery was simply assumed as one-tenth of the diameter $2 r_0$, i.e., $h_0 = r_0/5$.
Thus, the order of magnitude of the calculated Young's modulus, referring to the values of $r_0$ in Table~\ref{tab:vessels}, was $O(E) = 10^{-1}$ MPa,
which is consistent with that of aortic elasticity in conscious dogs~\cite{Armentano1991}.
The Young's modulus of the bypass was set to be $100$ times larger than that obtained with the reference radius ($r_0 = 4.5$ mm),
i.e., $E_\mathrm{bypass} = 43.3$ MPa.
Given that a previous numerical analysis of ePTFE stent grafts used a Young's modulus of $55.2$ MPa~\cite{Kleinstreuer2008},
we set the same order of magnitude of the Young's modulus in the stent grafted region, i.e., $E_\mathrm{s} = 10$ MPa.

The governing equations~\eqref{eq:govern} can be written as an advection equation:
\begin{align}
  \partial_t \v{v} + \mathbf{J}(\v{v}) \cdot \partial_x \v{v} = \v{b} \  \to
  \partial_t \v{v} + \partial_x \v{F} = \v{b},
  \label{eq:govern_2}
\end{align}
where the variable $\v{v}$,
advection term $\v{F}$,
source term $\v{b}$,
and Jacobian $\v{J}$ are written as
\begin{align}
  \v{v} = 
  \begin{pmatrix}
  Q \\
  A
  \end{pmatrix}, \quad
  \v{b} = 
  \begin{pmatrix}
  -K_R \frac{Q}{A} \quad
  0
  \end{pmatrix}, \quad
  \v{F} = 
  \begin{pmatrix}
  \frac{Q^2}{A} + \frac{\beta}{3 \rho} A^{3/2} \quad
  Q
  \end{pmatrix}, \quad
  \mathbf{J}(\v{v}) =
  \begin{bmatrix}
  2 \frac{Q}{A} & -\left( \frac{Q}{A} \right)^2 + \frac{\beta}{2 \rho}A^{1/2} \\
  1 & 0
  \end{bmatrix}.
\end{align}
The flow and pulse wave are integrated over time by the Lax-Wendroff method with $\Delta t = 10^{-5}$ s from $t^n$ to $t^{n+1} = t^n + \Delta t$.
The precise descriptions are given in the Appendix~\S{A.2}.
\begin{table*}[h!]
  \begin{center}
    \caption{Nomenclature for parameters and variables.}
   \label{tab:parameters}
    \begin{tabular}{c l l c} \hline
      Symbol  &  Physical meaning & Value & Reference \\ \hline
      $A (x, t)$ & Cross-sectional area at $x$ and at time $t$ & &\\
      $U (x, t)$ & Average fluid velocity at $x$ & & \\
      $p (x, t)$ & Pressure at $x$ & & \\
      $p_\mathrm{ext}$ & External pressure & $0$ Pa & - \\
      $p_a$ & Maximal pressure &  $4333$ Pa & - \\
      $p_0$ & Base pressure & $10666$ Pa & - \\
      $T$  & Wave period & $1$ s & - \\
      $\rho$ & Blood density & $1.05 \times 10^3$ kg/m$^3$ & - \\
      $\mu$ & Blood viscosity & $4.5 \times 10^{-3}$ Pa$\cdot$s & - \\
      $\m{V}$ & Blood kinematic viscosity ($= \mu/\rho$) & $4.3 \times 10^{-6}$ m$^2$/s & - \\
      $\nu$  & Poisson ratio & $0.5$ & - \\      
      $\alpha$  & Coefficient of the velocity profile & $1$ & \cite{Sherwin2003} \\
      $K_R$ & Drag coefficient ($= 22 \pi \m{V}$)  & & \\
      $r_0 (x)$ & Vessel radius at $x$ & & \\
      $A_0 (x)$ & Vessel cross-sectional area ($= \pi r_0^2$) at $x$ & &  \\
      $E (x, k_i)$ & Young's modulus at $x$ & &  \\      
      $k_i$ ($i = 1\mathrm{-}3$) & Coefficients for $E$ & $k_1 = 2 \times 10^6$ kg/(s$^2\cdot$m) & \cite{Olufsen1999, Smith2002} \\
      & & $k_2 = -2.253 \times 10^3$ m$^{-1}$ & \cite{Olufsen1999, Smith2002} \\
      & & $k_3 = 8.65 \times 10^4$ kg/(s$^2\cdot$m) & \cite{Olufsen1999, Smith2002} \\
      \hline
    \end{tabular}
  \end{center}
\end{table*}

\begin{figure}[htbp!]
  \begin{center}
  \includegraphics[width=8cm]{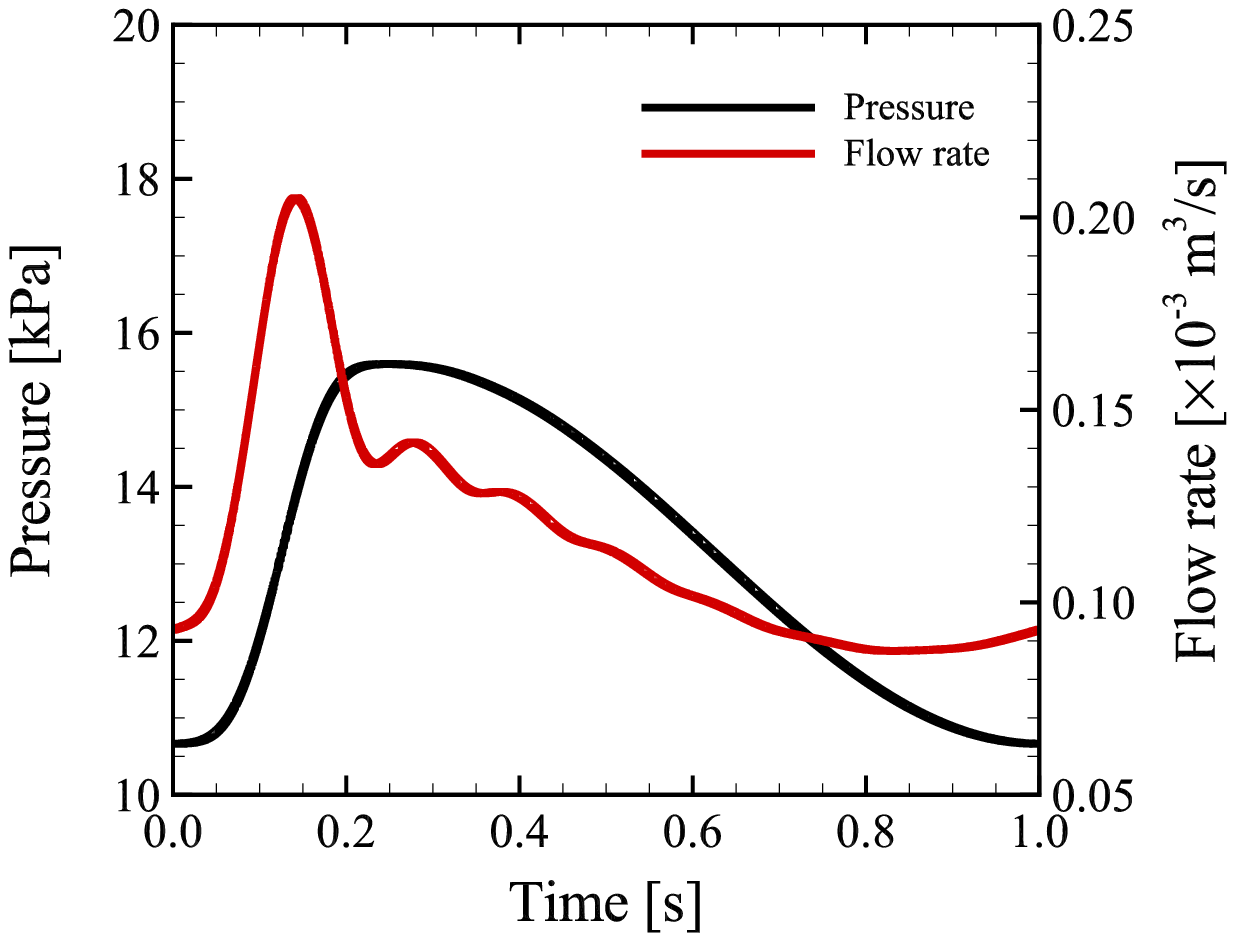}
  \end{center}
  \caption{The waveform of a pulsatile inlet pressure (black line) as a Dirichlet boundary condition,
  and the calculated inlet flow rate (red line) during a cardiac cycle $T$ ($= 1$ s).}
  \label{fig:input}
\end{figure}

To capture the internal pressure profile in the heart during a cardiac cycle,
wherein vascular pumping is accelerated in the systolic phase and attenuated in the diastolic phase~\cite{Morishita2021},
the inlet pressure $p_\mathrm{in}(t)$ is given as~\cite{Yokoyama2021}
\begin{align}
  p_\mathrm{in}(t) = p_0 +
  \begin{cases}
  \dfrac{p_a}{2}
   \left[1
   + \dfrac{1}{1-\epsilon_{\mathrm{s}}}
      \left(
        \sin\left(\pi \tilde{T}_{\mathrm{s}} \right)
     + \epsilon_{\mathrm{s}} \sin\left(3 \pi \tilde{T}_{\mathrm{s}} \right)
      \right)
  \right] \quad \mathrm{for} \quad
  0 \leq t  < \dfrac{T}{4}
  \\
  \dfrac{p_a}{2}
  \left[1
  - \dfrac{1}{1-\epsilon_\mathrm{d}}
    \left(
      \sin\left(\pi \tilde{T}_\mathrm{d} \right)
   + \epsilon_{\mathrm{d}} \sin\left(3\pi \tilde{T}_\mathrm{d} \right)
    \right)
  \right] \quad \mathrm{for} \quad
  \dfrac{T}{4} \leq t < T
  \end{cases}
,
\label{eq:kinematics}
\end{align}
and
\begin{align}
\tilde{T}_\mathrm{s} = \dfrac{t - T/8}{T/4}, \quad
\tilde{T}_\mathrm{d} = \dfrac{t - 5T/8}{3T/4},
\label{eq:periodfraction}
\end{align}
where $p_a$ ($= 4333$ Pa) is the amplitude from the base pressure $p_0$ ($= 10666$ Pa),
$T$ ($= 1$ s) is the wave period,
and $\epsilon_\mathrm{s}$ or $\epsilon_\mathrm{d}$ is the waveform parameter (black line in Figure~\ref{fig:input}).
Hereafter, the subscripts s and d denote the systolic and diastolic phases, respectively.
In this study, $\epsilon_\mathrm{s} = 0.1$ and $\epsilon_\mathrm{d} = 0$~\cite{Yokoyama2021} were used,
so that the derivatives up to the fourth of $p(t)$ are continuous.
The waveform reflects a fast expansion during the systolic phase $0 \leq t < T/4$ and a slow contraction during the diastolic phase $T/4 \leq t < T$.
The parameters in our simulations and their values are shown in Table~\ref{tab:parameters}.

Figure~\ref{fig:input} shows pressures and calculated flow rates during a period $T$ ($= 1$ s).
The model pressure given in equations equation~\eqref{eq:kinematics} and \eqref{eq:periodfraction} well captures the profile of human aortic pressure (Figure $3$a in the work by~\cite{vandeVosse2011}).
A no-reflection condition was applied for each outlet by a backward Riemann invariant $W_- = 0$.
A more precise description of the methodology is presented in the Appendix~\S{A.2}.

\section{Results}

\subsection{Measurements of flow rate in pre-and postoperative patients}
\begin{figure}[htbp!]
  \begin{center}
  \includegraphics[width=8cm]{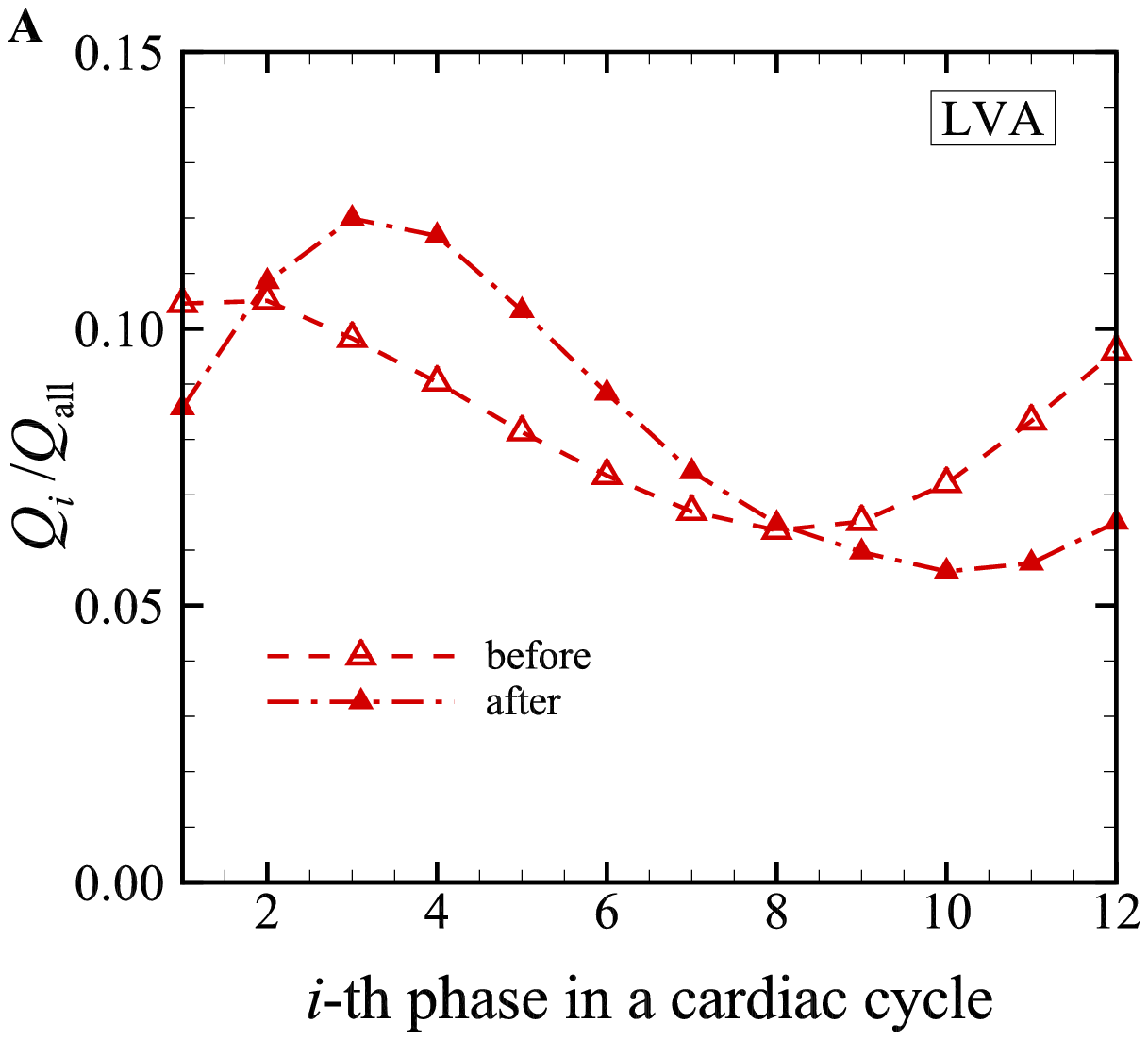}
  \includegraphics[width=8cm]{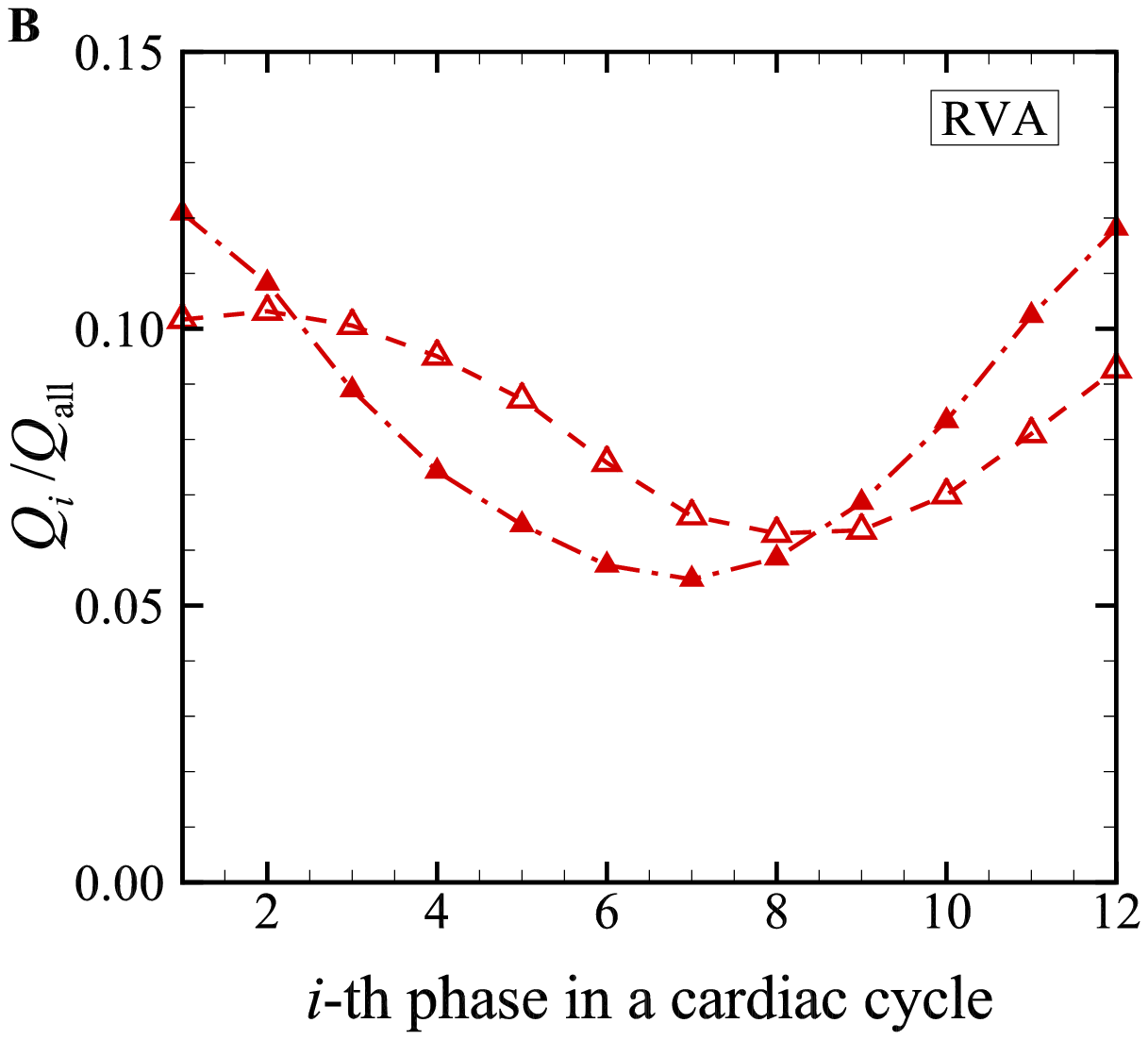}
  \end{center}
  \caption{Experimental measurements of the flow rate ratio $Q_i/Q_\mathrm{all}$ at $12$ phases in \textbf{(A)} the LVA and \textbf{(B)} the RVA during one cardiac cycle.
  The results were normalized by the total flux $Q_\mathrm{all}$ in a cardiac cycle $T$ ($= 1$ s).}
  \label{fig:exp}
\end{figure}

The flow rates in both the L/RVA in $9$ patients before and after $1$dTEVAR during a single heartbeat were measured by $2$D cine phase-contrast MRI.
The mean flow rates are shown in Figure~\ref{fig:exp},
where the heartbeat was divided into $12$ parts and the flow rate at the $i$-th phase $Q_i$ was normalized by the total flow rate $Q_\mathrm{all} (= \sum_i^{12} Q_i)$ in both the L/RVA during a period $T$.
The first phase (i.e., $1/12$) was defined when the pulse wave was detected by an accelerometer (GE Healthcare) with the tip of a forefinger.
Compared to the peak in the second phase of the cardiac cycle (i.e., $2/12$) in the LVA of preoperative patients,
the peak in postoperative patients was delayed to the third phase (i.e., $3/12$).
Thus, a $1/12$-period ($\sim8$\%) phase delay in the LVA was observed between before and after surgery (Figure~\ref{fig:exp}\textbf{(A)}).
On the other hand, 
the peak of the flow rate ratio in the RVA (the first phase, $1/12$) remained almost the same before and after surgery (Figure~\ref{fig:exp}\textbf{(B)}).
Since the difference of flow rate ratio in the RVA between the first and second phases of the cardiac cycle was very small,
we concluded that the first phase of the cardiac cycle was the time point of the peak of the flow rate ratio.

\subsection{Model validation}
\begin{figure*}[htbp!]
  \begin{center}
  \includegraphics[width=17cm]{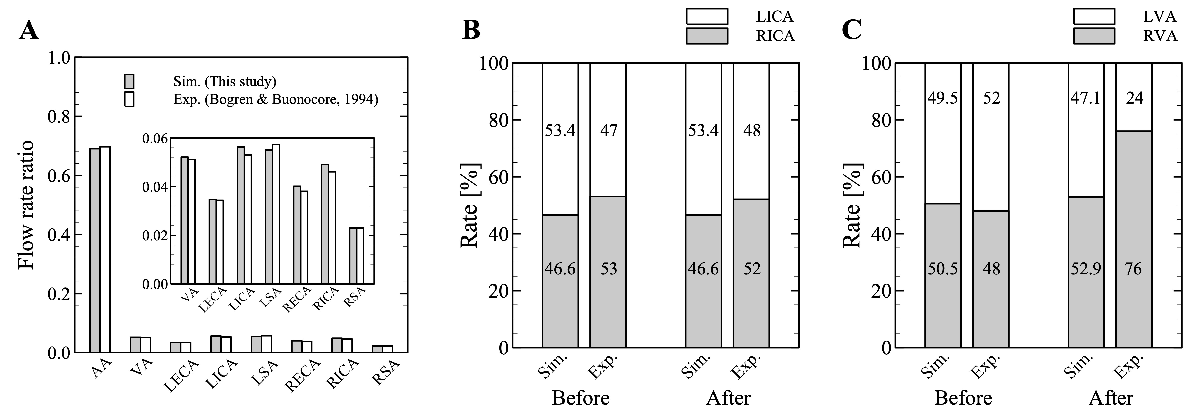}
  \end{center}
  \caption{
  \textbf{(A)} Comparison of the time-average flow rate ratio (AA, VA, L/RECA, L/RICA, and L/RSA) for the mean inlet flow rate $\bar{Q}_\mathrm{in} = (1/T) \int_0^T Q_\mathrm{in} dt$ between numerical results and experimental measurements by Bogren \& Buonocore~\cite{Bogren1994}.
   The results were obtained with the preoperative vascular tree model (see Figure~\ref{fig:schematics}\textbf{(A)})
   \textbf{(B)} Comparison of the time-average flow rate ratio of the L/RICA and 
   \textbf{(C)} the L/RVA between numerical results and experimental measurements by Goto et al.~\cite{Goto2019}.}
  \label{fig:comparison}
\end{figure*}

Using a preoperative vascular tree model (Figure~\ref{fig:schematics}),
the mean flow rates $Q_\mathrm{mean}$ in eight different arteries were investigated,
specifically AA, VA ($=$ LVA $+$ RVA), L/RICA, and L/RSA.
The calculated mean flow rates were normalized by the mean inlet flow rate $\bar{Q}_\mathrm{in} = (1/T)\int_0^T Q_\mathrm{in} dt$,
and the flow rate ratio $Q_\mathrm{RATIO} = Q_\mathrm{mean}/\bar{Q}_\mathrm{in}$ in each vessel was compared with data in a previous experimental study by Bogren \& Buonocore~\cite{Bogren1994} (Figure~\ref{fig:comparison}\textbf{A}).
The outlet diameters were determined so that the errors in the flow rate ratio in each vessel were uniformly less than $5$\%, i.e. $| Q_\mathrm{RATIO}^\mathrm{sim}/Q_\mathrm{RATIO}^\mathrm{exp} - 1| < 0.05$,
where the superscripts ``sim'' and ``exp'' denote the simulation and experiment, respectively. 

Using the same preoperative vascular tree model,
the flow rate ratio between the L/RICA and L/RVA was calculated.
The flow rate ratio in patients after $1$dTEVAR was also calculated with the postoperative vascular tree model (Figure~\ref{fig:schematics}\textbf{B}).
The diameters of the end terminals were the same as those in the preoperative vascular tree model.
The calculated flow rate ratio in each artery was compared with previous experimental measurements by Goto et al.~\cite{Goto2019} (Figures~\ref{fig:comparison}\textbf{B} and \ref{fig:comparison}\textbf{C}).
In the experiments,
although there were no large differences before and after surgery in the flow rate ratio in the L/RICA or the LVA,
the flow rate ratio in the RVA increased after surgery.
The numerical results qualitatively agree with these experimental measurements.

\subsection{$1$D model analysis of the flow rate in the ICA in pre-and postoperative patients}
\begin{figure}[htbp!]
  \begin{center}
  \includegraphics[width=8cm]{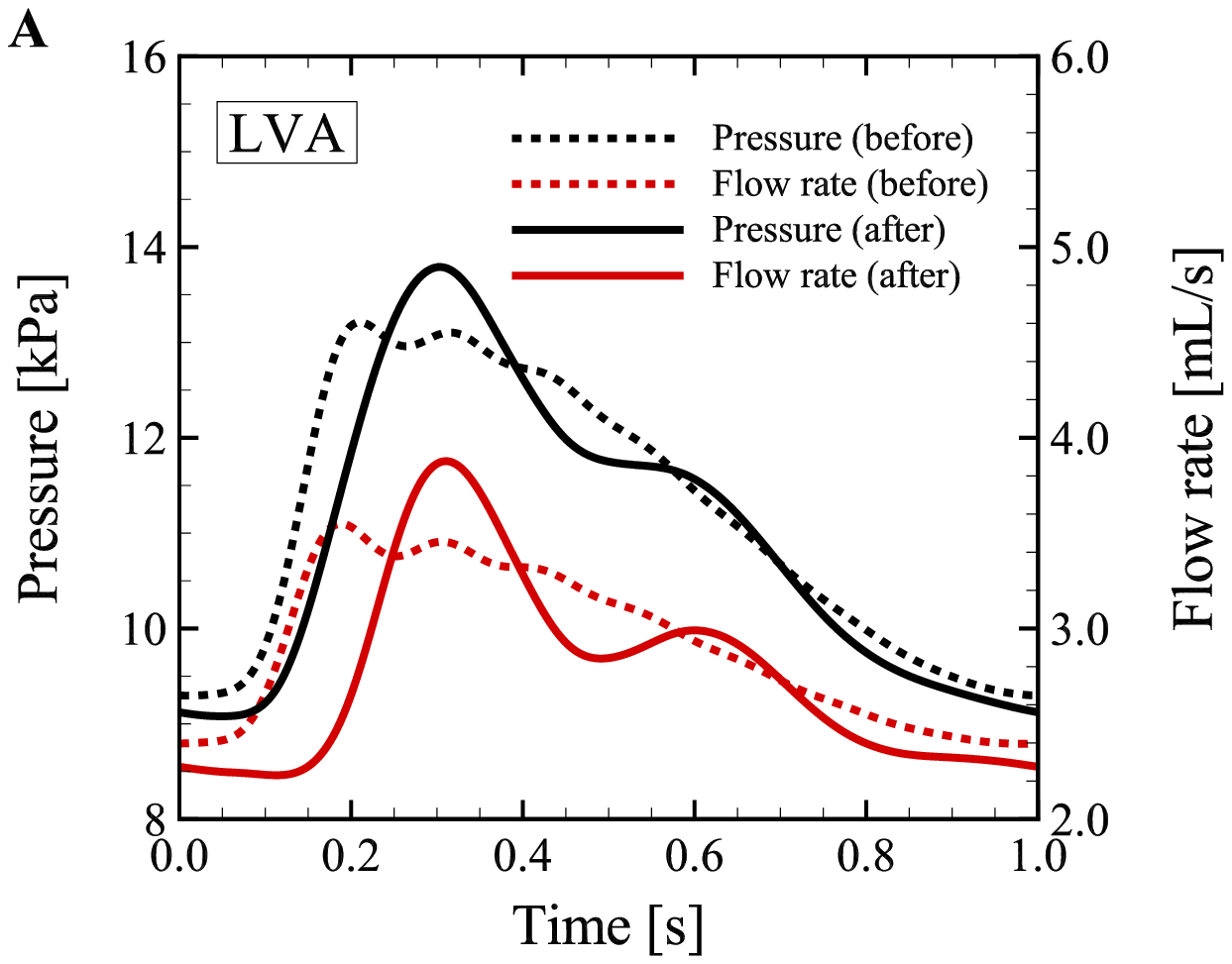}
  \includegraphics[width=8cm]{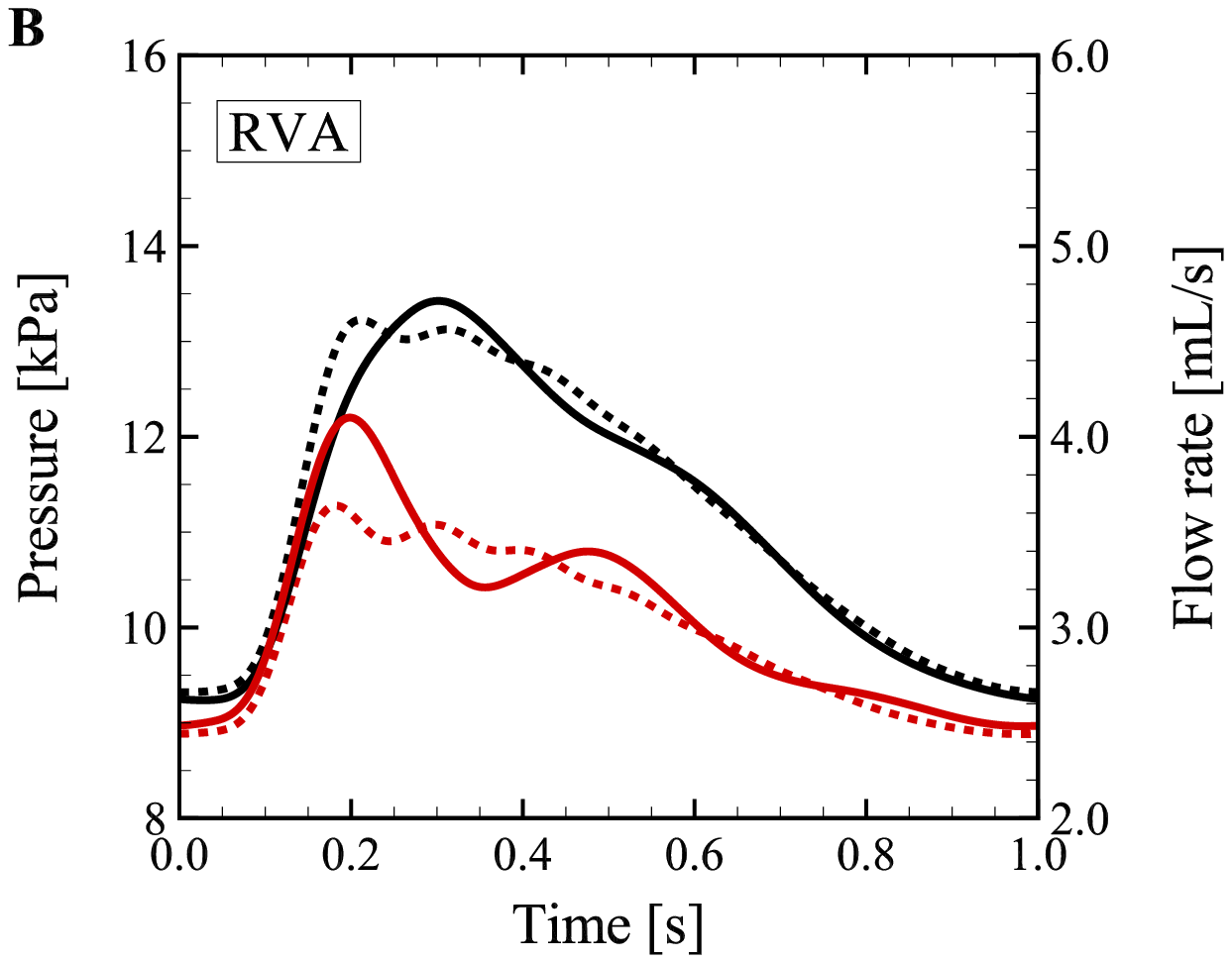}
  \end{center}
  \caption{The time histories of the calculated pressure $p$ [kPa] (left axis) and flow rate $Q$ [mL/s $= 10^{-6}$ m$^3$/s] (right axis) in \textbf{(A)} the LVA and \textbf{(B)} the RVA for pre-and postoperative models (dashed and solid lines, respectively) during a cardiac cycle $T$,
  where data are shown after a two-period ($2$-second) pass.}
  \label{fig:results_pre_post}
\end{figure}

To gain insight into the mechanism of the phase difference in the LVA and the lack of such a difference in the RVA (Figure~\ref{fig:exp}),
$1$D numerical simulations were performed using arterial vascular tree models before and after surgery (Figure~\ref{fig:schematics}).
Model verifications are shown in the Appendix~\S{A.3} (see Figures~\ref{fig:verification}).
The time history of the pressure and flow rates in the L/RVA during a cardiac cycle $T$ are shown in Figure~\ref{fig:results_pre_post}\textbf{(A)} and \ref{fig:results_pre_post}\textbf{(B)}, respectively,
where the results before and after surgery are superposed.
Data are shown after the pressure and flow rates have reached the stable periodic phase ($t \geq 2$ s).
As described regarding the experimental measurements of flow rates (Figure~\ref{fig:exp}\textbf{(A)}),
the numerical results of the flow rate waveform in the LVA were delayed, by $6.8$\% of a single period,
which is similar to the experimental measurements (Figure~\ref{fig:results_pre_post}\textbf{(A)}).
As in experimental measurements, the phase difference was also quantified by the time point observed at the maximum flow rate.
On the other hand,
the numerical results of the flow rate waveform in the RVA were only slightly delayed, by $0.84$\% of a single period (Figure~\ref{fig:results_pre_post}\textbf{(B)}).
We concluded that our numerical results of the phase difference in the L/RVA are consistent with the experimental measurements (Figure~\ref{fig:exp}\textbf{(B)}).

\subsection{\label{effects_stent_stiffness}Effects of local vessel stiffness and flow path}
\begin{figure*}[htbp!]
  \begin{center}
  \includegraphics[width=8cm]{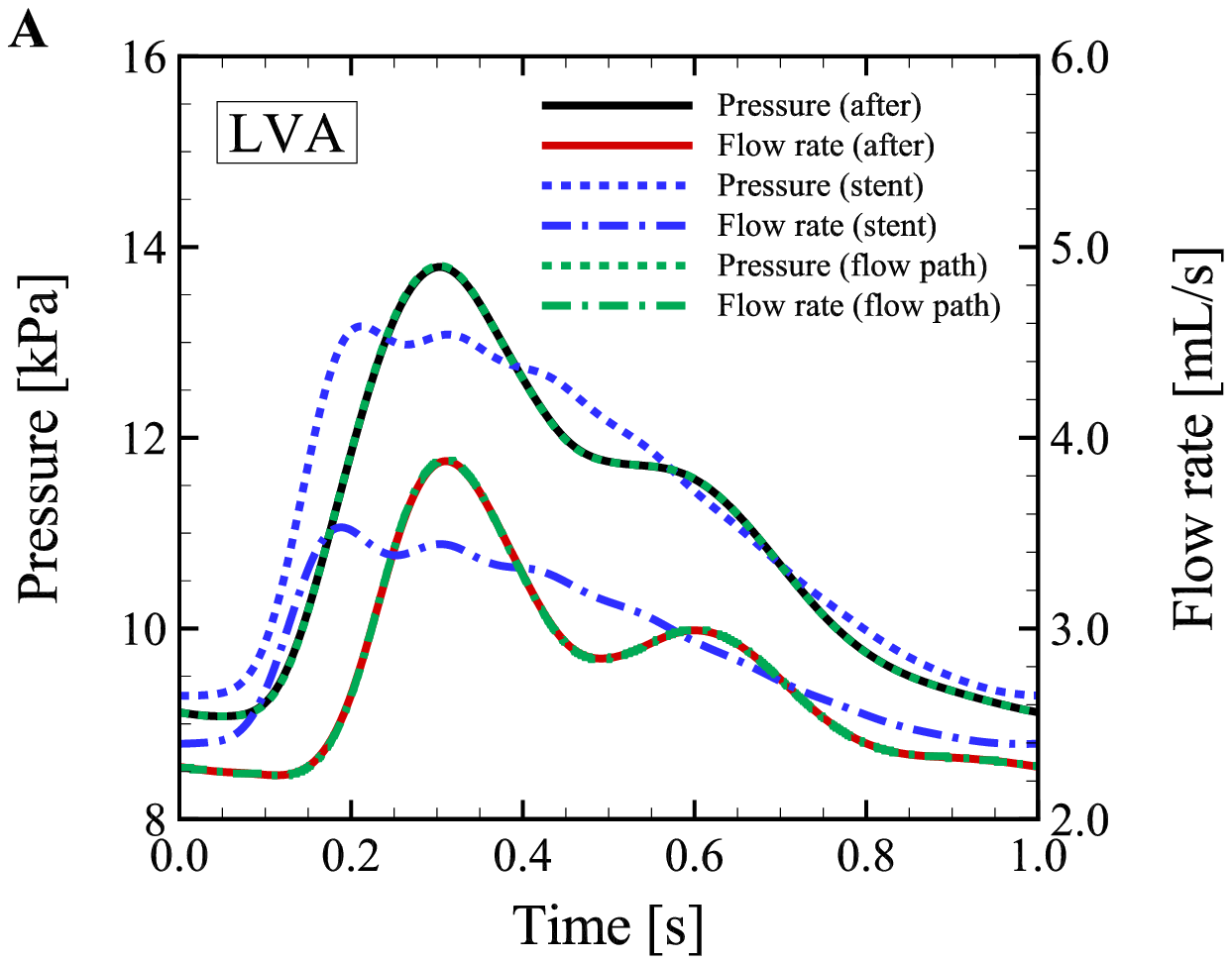}
  \includegraphics[width=8cm]{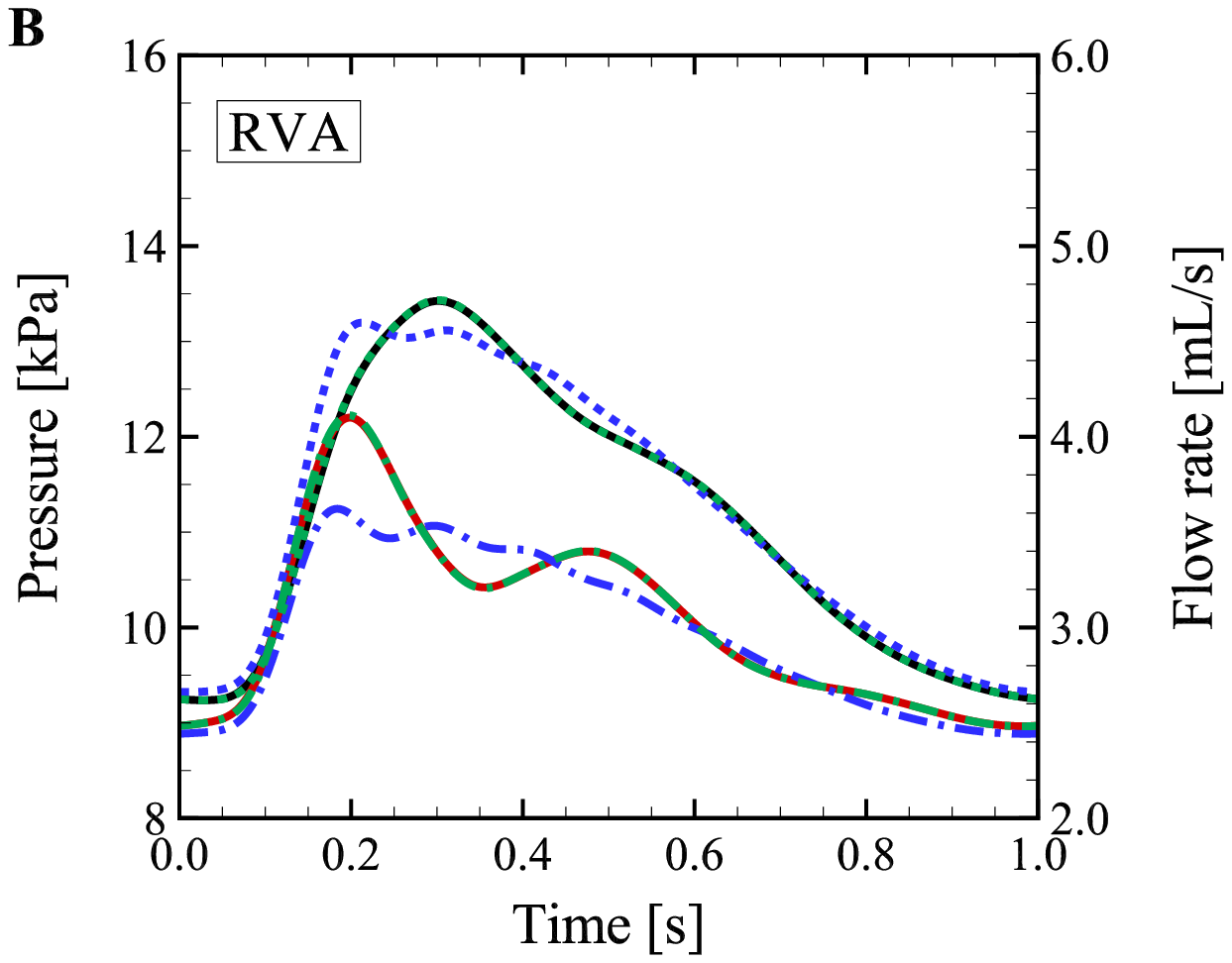}
  \includegraphics[width=8cm]{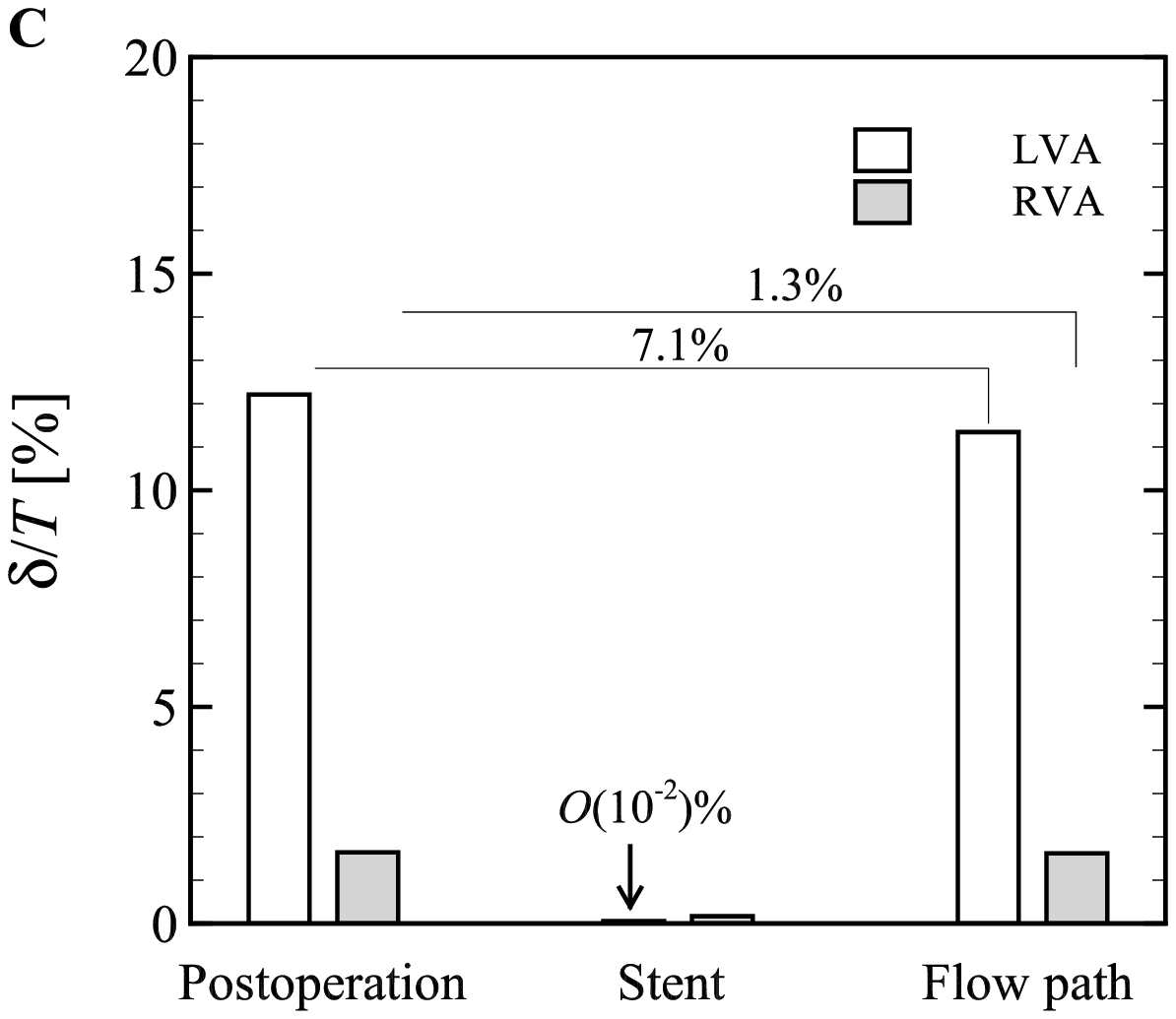}
  \end{center}
  \caption{The time histories of the calculated pressure $p$ and flow rates $Q$ in \textbf{(A)} the LVA and \textbf{(B)} the RVA during a cardiac cycle $T$ ($= 1$ s),
  where data are shown after a two-period pass.
  ``Stent'' denotes the results obtained with the vascular tree model that increases the Young's modulus in the stent-grafted area.
  ``Flow path'' denotes the results obtained with the vascular model that changes the flow path in the same way as the postoperative vascular model (i.e., including the bypass and embolization, see Figure~\ref{fig:schematics}\textbf{(B)}) while retaining the same Young's modulus as the preoperative model.
  \textbf{(C)} The ratio of the phase difference of the flow rate $\delta/T$ in both the L/RVA between the preoperative vascular tree model and the two aforementioned models (stent and flow path).
  The results were normalized by $T$.}
  \label{fig:results_effects}
\end{figure*}

To clarify whether the flow rate waveforms in the LVA (Figure~\ref{fig:results_pre_post}\textbf{(A)}) can be caused by the local vessel stiffness, flow path changes, or both,
the effect of both of these factors on the flow rate in the L/RVA was investigated with two different additional vascular tree models.
The Young's modulus in the stent-grafted region was increased by $100$ times postoperatively compared to the preoperative state without the bypass or embolization, the so-called ``stent model",
while the flow path change was the same as that in the postoperative vascular tree model (i.e., new bypass and embolization) and the Young's modulus was the same as that in the preoperative model, the so-called ``flow path model".
The flow rate and pressure waveforms during a cardiac cycle $T$ are shown in Figure~\ref{fig:results_effects}\textbf{(A)} and \ref{fig:results_effects}\textbf{(B)}.
The flow rate waveforms obtained with the stent model were different from those in the postoperative vascular tree model in both the L/RVA.
On the other hand,
the flow rate waveforms obtained with the flow path model collapsed with those obtained with the preoperative vascular tree model.
These agreements or disagreements were also commonly observed in the pressure waveforms.
The phase difference $\delta$ of the maximum flow rate in both the L/RVA between the preoperative vascular tree model and the three aforementioned models (stent, flow path, and postoperative) are summarized in Figure~\ref{fig:results_effects}\textbf{(C)},
where the results were normalized by $T$.
Changing the flow path, as in the flow path model, decreased the ratio of the phase difference of the flow rate $\delta/T$ by only $7.1$\% in the LVA and $1.3$\% in the RVA relative to the ratios obtained in the postoperative model.
On the other hand, changing the local Young's modulus, as in the stent model, caused quite small phase differences, i.e., $O(\delta/T) = 10^{-2}\%$.

\section{Discussion}\label{discussion}
A previous clinical study by Goto et al.~\cite{Goto2019} showed hemodynamic changes in blood flow through the L/RVA after $1$dTEVAR,
while postoperative total intracranial blood flow was almost the same as that measured preoperatively.
This might be due to systemic hemodynamic compensation causing cerebral blood flow to be strictly maintained through cerebral autoregulation~\cite{Seifert2011}.
Despite the maintained flow mass in patients after $1$dTEVAR,
experimental measurements demonstrated phase differences of the flow rate in the LVA but not in the RVA between pre- and postoperative patients undergoing $1$dTEVAR (Figure~\ref{fig:exp}).
To the best of our knowledge,
this study is the first report to describe phase differences of the flow rate in LVAs after 1dTEVAR with AxA-AxA bypass.
It is expected that the aforementioned hemodynamic compensation and the phase differences in the VA in postoperative patients might be due to two mechanical factors:
structural alterations in flow paths due to the new bypass and embolization, and an increase of local vessel stiffness due to stent grafting.
However, much is still unknown about this matter,
in particular about the relative contribution of these two factors to the phase of the flow rate in the VA between pre- and postoperative patients undergoing $1$dTEVAR.
To explore this issue,
A $1$D model was used to numerically investigate flow rates in the VA, both before and after $1$dTEVAR.
Using a postoperative arterial tree model (Figure~\ref{fig:schematics}\textbf{(B)}),
The numerical results demonstrated a phase delay in the LVA compared to preoperative models, as shown in Figure~\ref{fig:results_pre_post}\textbf{(A)}.
Stent grafting,
which is characterized by a locally increasing Young's modulus,
contributed negligibly to the phase delay in the LVA (see \S\ref{effects_stent_stiffness}).
Thus, the experimentally observed phase delay (Figure~\ref{fig:exp}\textbf{(A)}) was mainly caused by alteration of the flow path, i.e., by the new bypass and embolization, rather by local vessel stiffness due to stent grafting (Figure~\ref{fig:results_effects}).

As reported in previous study~\cite{Goto2019},
the 30-day mortality and inpatient death rates were both 0\% in patients after $1$dTEVAR.
Furthermore, no postoperative symptomatic stroke was observed in postoperative patients.
Thus, a phase delay in the flow rate of the LVA after 1dTEVAR, relative to before, does not affect the postoperative stroke,
at least in the short-term postoperative prognosis.
Although further statistical analysis is required due to the small number of enrolled patients,
at the present, there is no clinical evidence that postoperative phase delay may be critical to their quality of life (QOL).
A follow-up study is needed to clarify whether the postoperative phase delay in LVA can be related to long-term cerebral complications or prognosis.
To evaluate the long-term impact of such a significant phase difference on posterior cerebral circulation,
follow-up MRI should be performed periodically,
which is however our future study.


In general, the morphometric properties of the aorta differ along its length.
In healthy humans,
helical computed tomography has showed that the maximal aortic diameter is located in the ascending aorta, just distal to the aortic valve sinus and proximal to the innominate artery~\cite{Hager2002}.
The aortic diameter then progressively decreases along the thoracic aorta and continues to decrease from the infrarenal abdominal aorta to the lower abdominal aorta~\cite{Hager2002, Rogers2013}.
Similarly, the overall thickness of the aortic wall also decreases along the thoracic aorta but remains constant in the abdominal aorta~\cite{Sokolis2007}.
These findings are qualitatively similar to our measurements of reference radius $r_0$ in Table~\ref{tab:vessels} and consistent with our modeling (equations\eqref{eq:kinematics} and \eqref{eq:periodfraction}).
It is also known that aging causes morphometric changes in the aforementioned geometric parameters of the aorta (diameter, length, and thickness)~\cite{Hickson2010}.
Additionally, aortic stiffness increases significantly with age and is associated with an elevated risk for various adverse clinical outcomes,
e.g., heart disease, dementia, and kidney disease~\cite{Mitchell2021}.
In this study,
the geometrical parameters in Table~\ref{tab:vessels} and mechanical property represented by Young's modulus referred to normal subjects.
Given that preoperative patients have the aortic aneurysm and have potential complications,
selecting these parameters for preoperative patients will lead to a more rigorous estimation of the phase delay.
However, a systematic parametric study of these parameters is beyond the scope of the present work.

In this simulation,
branch angles were uniformly set to $30$ deg for simplicity.
A previous experimental study showed that the energy loss at bifurcations was very small as reported in~\cite{Matthys2007}.
This is because the branch angle only affects the total pressure continuity (equations~\ref{eq:3to1and2}(b, c) and \ref{eq:1and2to3}(b, c), see the Appendix~\S{A.1}),
and the order of magnitude of the static pressure term $p$ is much greater than those of the other terms.
Indeed,
the relative difference of the mean flow rate in the RVA obtained with the reference angle ($30$ deg) and that obtained with $120$ deg at the branch angle of the bypass is smaller than 1\% (data not shown).
Thus, at least in this model, the phase delay in the LVA was insensitive to the bypass angle $\theta$.
We did not consider postoperative shape deformation and diameter changes due to device placement.
In the future we will perform systematic analyses of the effect of postoperative vascular configurations on the phase change.

In this study, the Young's modulus in model vasculatures was also approximated using equation~\eqref{eq:E}.
Referring to previous numerical assessments of the mechanical behavior of ePTEF,
we set the same order of magnitude of the Young's modulus for the stent-grafted area (i.e., $E_\mathrm{s} =10$ MPa),
which was approximately 100 times larger than that of vessels with similar diameter.
Considering different Young's moduli of PET stent from $O(10^0)$ MPa~\cite{Kleinstreuer2008, Kan2021, Ramella2022} to $O(10^3)$ MPa~\cite{Ramella2022, Ma2018} in previous numerical studies,
the simulations were performed for different orders of magnitude of Young's modulus in an ePTFE-stent-grafted area ($E/E_\mathrm{s} = 0.01-10$).
However, the ratio of the phase difference $\delta/T$ remained almost the same (less than 0.2\%) in both the L/RVA.
Since model parameters in the bifurcation model ($\gamma_1$ and $\gamma_2$ in equations~\ref{eq:3to1and2}(b, c) and \ref{eq:1and2to3}(b, c), see the Appendix~\S{A.2}) and in Young's modulus ($k_1, k_2$, and $k_3$ in equation~\eqref{eq:E}) were fixed in this study,
further precise analysis of the effects of bifurcation angles and Young's modulus of the stent should be performed experimentally to confirm whether those mechanical factors have less impact on the phase differences.

The velocity profile was implicitly expressed as flat,
derived from Newtonian and laminar fluid flow.
However, it is expected that the blood flow profile, especially in a real artery, is much more complex,
and therefore it is usually modeled as plug flow due to its turbulent nature and cellular dynamics.
Since the Reynolds number in the aorta,
estimated based on the mean flux between systolic and diastolic phases,
is over 4 $\times$ 10$^3$,
the flow should be turbulent.
Furthermore, the blood velocity profile is also affected by frequency-dependent inertia.
The ratio between the transient inertia force and the viscous force can be estimated by the Womersley number $Wo = r_0 (\omega/\m{V})^{1/2} \approx 1.2-12$,
where $r_0$ ($= 1-10$ mm) is the radius of the artery (see Table~\ref{tab:vessels}),
$\omega$ ($= 2 \pi f$) is the angular frequency,
$f$ ($= 1/T$) is the cardiac frequency that is roughly estimated as $1$ Hz,
and $\m{V}$ ($= \mu/\rho \approx 4.3 \times 10^{-6}$ m$^2$/s) the kinematic viscosity of the blood.
Thus, it is expected that a more rigorous velocity flow profile can reproduce unsteady hemodynamics in patients before and after dTEVAR.
In this study, however, we focused on explaining the experimentally observed phase delay of the mean volumetric flow rate in the LVA seen in postoperative $1$dTEVAR patients relative to that before surgery,
and thus, the flow profile predominantly affecting the unit mean volumetric flow rate should not change the calculated phase delay.

In this simulation,
we simply modeled arterial trees from the heart (or LV) to eight end terminals,
whose diameters and lengths were determined so that the time average flow rate became similar to that obtained in experimental measurements by Bogren \& Buonocore~\cite{Bogren1994} or Goto et al.~\cite{Goto2019} (Figure~\ref{fig:comparison}).
Even with the simplest boundary condition at the end terminals (no-reflection condition, i.e., $W_- = 0$),
our numerical model reproduced the phase delay of the flow rate in the LVA without the phase difference in the RVA in postoperative patients,
and also clarified that the phase delay arose mostly from alteration of the flow path (Figure~\ref{fig:results_effects}).
Further precise boundary conditions for both the inlet (heart) and outlets will leads to patient-specific analyses preoperatively and to the evaluation of hemodynamics postoperatively,
topics that we will address in the future.

Depending on the areas of TAA,
different numbers of bypasses (1d, 2d, and 3d) can be selected.
If phase delays of flow rate are present in the LVA of postoperative patients who have undergone different surgical operations,
it is expected that the phase differences between before and after surgery are caused by the new bypasses.
Quantitative analyses of hemodynamic changes caused by these surgeries should be performed in future studies.
Modification of the aforementioned model factors, e.g., boundary conditions, will potentially reproduce hemodynamics in patients who undergo these various surgeries.
The numerical results based on pulse-wave dynamics provide fundamental knowledge regarding hemodynamic changes between pre- and postoperative patients undergoing dTEVAR,
and will be helpful not only in surgical decision-making for optimal clinical outcomes but also in evaluating hemodynamics after surgery.

\section{Conclusion}
This study is the first report to describe phase differences of the flow rate in LVAs after 1dTEVAR with AxA-AxA bypass.
A 1D model was used to numerically investigate blood flow rates,
with the goal of explaining experimental evidence on the phase delay of the flow rate in the LVA but not in the RVA after $1$dTEVAR relative to before.
The numerical model can reproduce the flow distribution in the major arteries from the heart,
and can capture the flow rate ratio in the L/RVA in both pre- and postoperative patients.
The numerical results showed that the phase delay was mainly caused by the bypass, i.e., by alteration of the flow path, rather than by stent grafting, i.e., the change of local vessel stiffness.
Bypass angles and the effects of the length and Young's modulus of the stent were also investigated,
but all were insensitive to the phase delay.
We hope that our numerical results will provide fundamental knowledge about therapeutic decisions for dTEVAR.


\section*{Ethical Approval}
Not required.

\section*{Declaration of Competing Interest}
The authors declare that they have no known competing financial interests or personal relationships that could have appeared to influence the work reported in this paper.

\section*{Acknowledgements}
The presented study was partially funded by Daicel Corporation.
Last but not least, N.T. and N.Y. thank Mr. Tatsuki Shimada for his assistance in the preparation of this work.

\appendix
\renewcommand{\thefigure}{A\alph{section}\arabic{figure}}
\renewcommand{\thetable}{A\alph{section}\arabic{table}}
\setcounter{figure}{0}

\section*{Appendix}
\subsection*{A1. Effects of stent grafting and bypass angle}
The effects of the local vessel stiffness (Young's modulus) and stent-grafted length on the phase difference in the L/RVA were 
investigated using the postoperative vascular tree model.
Simulations were performed for different Young's modulus valuses $E/E_\mathrm{s}$ ($= 0.01, 0.1, 1, 10$) in the stent-grafted artery with standard length $L_0 = 40$ mm,
where $E_\mathrm{s}$ ($= 10$ MPa) was the original Young's modulus in the stent-grafted region.
The ratio of the phase difference $\delta/T$ remained almost the same (less than $0.2$\%) in both the L/RVA.
Even when the length of the stent-grafted region increased to $L/L_0 = 1, 2.25, 3.5, 4.75, 6$,
the results of $\delta/T$ remained the same (less than $0.1$\%; data not shown),
where $L_0$ ($= 40$ mm) was the original length of the stent-grafted region.

\begin{figure}[htbp!]
  \begin{center}
  \includegraphics[width=8cm]{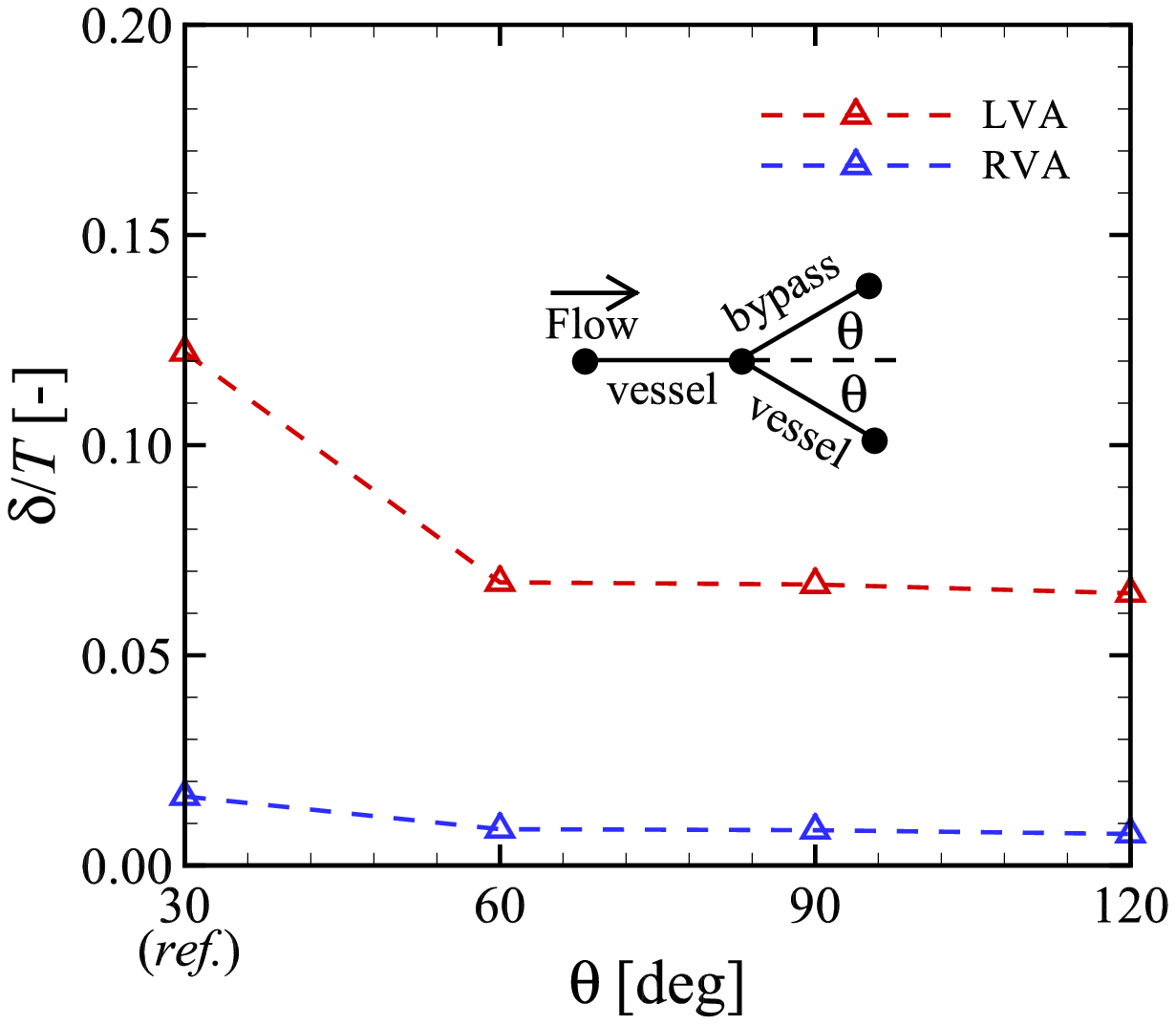}
  \end{center}
  \caption{Normalized phase difference of the flow rate $\delta/T$ in both the LVA and RVA between the preoperative and postoperative vascular tree models with different branch angles $\theta$ at the vessel-bypass.
  }
  \label{fig:results_angle}
\end{figure}

The phase delay in the LVA was also insensitive to the bypass angle $\theta$,
defined at the branch point between the vessel and bypass,
especially for $\theta \geq 60$ deg as shown in Figure~\ref{fig:results_angle}.
Note that the relative difference of the mean flow rate in the RVA obtained with the reference angle ($30$ deg) and that obtained with $120$ deg at the branch angle of the bypass was smaller than 1\% (data not shown).
The results were obtained with the postoperative vascular tree model with a standard stent length of $L_0 = 40$ mm and a Young's modulus of $E_\mathrm{s} = 10$ MPa.
These results, including those shown in Figure~\ref{fig:results_effects}\textbf{(C)}, suggest that the phase delay of the flow rate in the VA between before and after surgery (Figure~\ref{fig:exp}) arises mostly from the alteration of the flow path, i.e., by the new bypass and embolization rather than by local vessel stiffness due to stent grafting.

Note that the effect of the stiffness (or Young's modulus $E$) of the straight tube on the fluid velocity is described in \S{A.3}.
and the results showed that the pulse-wave speed $c$ increased with $E$ with the ratio of velocity $O(c/U) > 10^2$ (Figure~\ref{fig:verification}\textbf{D}).

\subsection*{A2. Methodology}
The variables of the $i$th-vessel segment at specific time $t^n$ ($\v{v}_i^n$ and $\v{F}_i^n$) in equation~\eqref{eq:govern_2} are described as:
\begin{equation}
  \v{v}_i^n = 
  \begin{pmatrix}
  Q_i^n \\
  A_i^n
  \end{pmatrix}, \quad
  \v{F}_i^n =
  \begin{pmatrix}
  \dfrac{(Q_i^n)^2}{A_i^n} + \dfrac{\beta}{3 \rho} (A_i^n)^{3/2} \\
  Q_i^n
  \end{pmatrix}.
\end{equation}
The discretized form of equation~\eqref{eq:govern_2} is obtained with the Lax-Wendroff method:
\begin{align}
  \v{v}_i^{n+1} 
  &= \v{v}_i^n
  - \frac{1}{2 \Delta c} \left( \v{F}_{i+1}^n - \v{F}_{i-1}^n \right) \nonumber \\
  &\hspace{0.5cm} + \frac{1}{2 (\Delta c)^2} \left[ \mathbf{J} \left( \frac{\v{v}_i^n + \v{v}_{i+1}^n}{2} \right) \left( \v{F}_{i+1}^n - \v{F}_i^n \right)
  - \mathbf{J} \left( \frac{\v{v}_i^n + \v{v}_{i-1}^n}{2} \right) \left( \v{F}_i^n - \v{F}_{i-1}^n \right) \right], \\
  &= \v{v}_i + \frac{1}{2 \Delta c} \left[ \v{F}_{i-1} + \v{F}_i^n
  + \frac{1}{\Delta c} \mathbf{J} \left( \frac{\v{v}_i^n + \v{v}_{i-1}^n}{2} \right) \left( \v{F}_{i-1}^n - \v{F}_i^n \right) \right. \nonumber \\
  &\hspace{1.8cm} \left. - \left\{ \v{F}_i^n + \v{F}_{i+1}^n + \frac{1}{\Delta c} \mathbf{J} \left( \frac{\v{v}_i^n + \v{v}_{i+1}^n}{2} \right) \left( \v{F}_i^n - \v{F}_{i+1}^n \right) \right\} \right],
\end{align}
where $\Delta c = \Delta x/\Delta t$, and $\Delta x$ is the segment length of the vessel.

The boundary values of $Q$ and $A$ are determined by the Riemann invariants ($W_+$ and $W_-$),
which represent a forward- and backward-traveling wave at speeds $\lambda_+$ and $\lambda_-$ as eigen values of the Jacobian $\mathbf{J}$.
Riemann invariants are the characteristic variables of the following hyperbolic system transformed from equation~\eqref{eq:govern_2}:
\begin{align}
  &\partial_t \v{W} + \v{\lambda} \cdot \partial_x \v{W} = 0,
  \label{eq:hyperbolic_system}
\end{align}
where $\v{W} = (W_+, W_-)^T$,
and $\v{\lambda} = (\lambda_+, \lambda_-)^T$.
By choosing the reference conditions ($A = A_0$, $U = 0$),
we obtain the solutions to system~\eqref{eq:hyperbolic_system}:
\begin{align}
  W_\pm
  &= \frac{Q}{A} \pm 4 \sqrt{\frac{\beta}{2 \rho}} \left( A^{1/4} - A_0^{1/4} \right), \label{eq:W_pm} \\
  &= U \pm 4 \left( c - c_0\right), \label{eq:Riemann}\\
  \text{and} \quad
  \lambda_\pm &= U \pm c, \label{eq:lam_pm}
\end{align}
where $c = \sqrt{\beta/(2 \rho)} A^{1/4}$ is the wave speed,
and $c_0$ is the wave speed at $A = A_0$.

\begin{figure}[htbp!]
  \begin{center}
  \includegraphics[width=10cm]{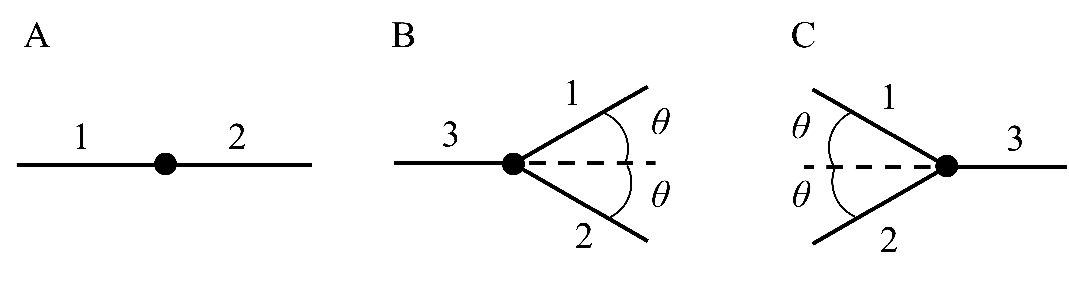}
  \end{center}
  \caption{
  Three different types of flow conditions:
  \textbf{(A)} a simple connection from segment $1$ to segment $2$ ($1 \to 2$),
  \textbf{(B)} a bifurcation ($3 \to 1 + 2$), and
  \textbf{(C)} a confluence ($1 + 2 \to 3$),
  where $\alpha$ is the branch angle.
  }
  \label{fig:connection}
\end{figure}
Flow conditions along the lines ($1 \to 2$) (see Figure~\ref{fig:connection}\textbf{(A)}) are described by mass conservation and total pressure continuity~\cite{Formaggia2003}:
\begin{subequations}
\begin{align}
  &Q_1 - Q_2 = 0, \\
  &p_1 + \rho U_1^2/2 - \left( p_2 + \rho U_2^2/2 \right) = 0,
\end{align}
\end{subequations}
and $W_1^{n+1}(L)$ and $W_2^{n+1}(0)$ are derived from values at the previous time step at the distal end (denoted by $L$) and proximal end (denoted by 0) of an artery by extrapolating the outgoing Riemann invariants along the characteristic lines~\cite{Formaggia2003}:
\begin{subequations}
\begin{align}
  &W_{1, +}^{n+1} (L) = W_{1, +}^{n} \left( L - \lambda_{1, +} \Delta t \right), \\
  &W_{2, -}^{n+1} (0) = W_{2, +}^{n} \left( 0 - \lambda_{2, -} \Delta t \right).
\end{align}
\end{subequations}
Flow conditions at the bifurcations ($3 \to 1 + 2$) (Figure~\ref{fig:connection}\textbf{(B)}) are described as~\cite{Formaggia2003}:
\begin{subequations}
\begin{align}
  &-Q_1 - Q_2 + Q_3 = 0, \\
  &p_3 + \rho U_3^2/2 - \mathrm{sign}(Q_3) f_\mathrm{m} (Q_3/A_3) - \left[ p_1  + \rho U_1^2/2 + \mathrm{sign}(Q_1) f_\mathrm{t} \left( Q_1\middle/\frac{A_1 + A_3}{2} \right) \right] = 0, \\
  &p_3 + \rho U_3^2/2 - \mathrm{sign}(Q_3) f_\mathrm{m} (Q_3/A_3) - \left[ p_2  + \rho U_2^2/2 + \mathrm{sign}(Q_2) f_\mathrm{t} \left( Q_2\middle/\frac{A_2 + A_3}{2} \right) \right] = 0, \\
  &W_{3, +}^{n+1} (L) - W_{3, +}^n (L - \lambda_{3, +} \Delta t) = 0, \\
  &W_{1, -}^{n+1} (0) - W_{1, -}^n (0 - \lambda_{1, -} \Delta t) = 0, \\
  &W_{2, -}^{n+1} (0) - W_{2, -}^n (0 - \lambda_{2, -} \Delta t) = 0,
\end{align}
\label{eq:3to1and2}
\end{subequations}
where
\begin{subequations}
\begin{align}
  &f_\mathrm{m} (Q/A) = \gamma_1 \rho (Q/A)^2, \\
  &f_\mathrm{t} (Q/A) = \gamma_2 \rho (Q/A)^2 \sqrt{2 \left( 1 - \cos\alpha \right)}.
\end{align}
\end{subequations}
In this study, we set $\gamma_1 = 0$, $\gamma_2 = 2$, and $\alpha = \pi/6$.
At the confluences ($1 + 2 \to 3$) (Figure~\ref{fig:connection}\textbf{(C)}), we have:
\begin{subequations}
\begin{align}
  &Q_1 + Q_2 - Q_3 = 0, \\
  &p_1 + \rho U_1^2/2 - \mathrm{sign}(Q_1) f_\mathrm{t} \left( Q_1\middle/\frac{A_1 + A_3}{2} \right) - \left[ p_3  + \rho U_3^2/2 + \mathrm{sign}(Q_3) f_\mathrm{m} (Q_3/A_3) \right] = 0, \\
  &p_2 + \rho U_2^2/2 - \mathrm{sign}(Q_2) f_\mathrm{t} \left( Q_2\middle/\frac{A_2 + A_3}{2} \right) - \left[ p_3  + \rho U_3^2/2 + \mathrm{sign}(Q_3) f_\mathrm{m} (Q_3/A_3) \right] = 0, \\
  &W_{1, +}^{n+1} (L) - W_{1, +}^n (L - \lambda_{1, +} \Delta t) = 0, \\
  &W_{2, +}^{n+1} (L) - W_{2, +}^n (L - \lambda_{2, +} \Delta t) = 0, \\
  &W_{3, -}^{n+1} (0) - W_{3, +}^n (0 - \lambda_{3, -} \Delta t) = 0.
\end{align}
\label{eq:1and2to3}
\end{subequations}

\subsection*{A3. Weakly non-linear form}
\begin{figure*}[htbp!]
  \begin{center}
  \includegraphics[width=8cm]{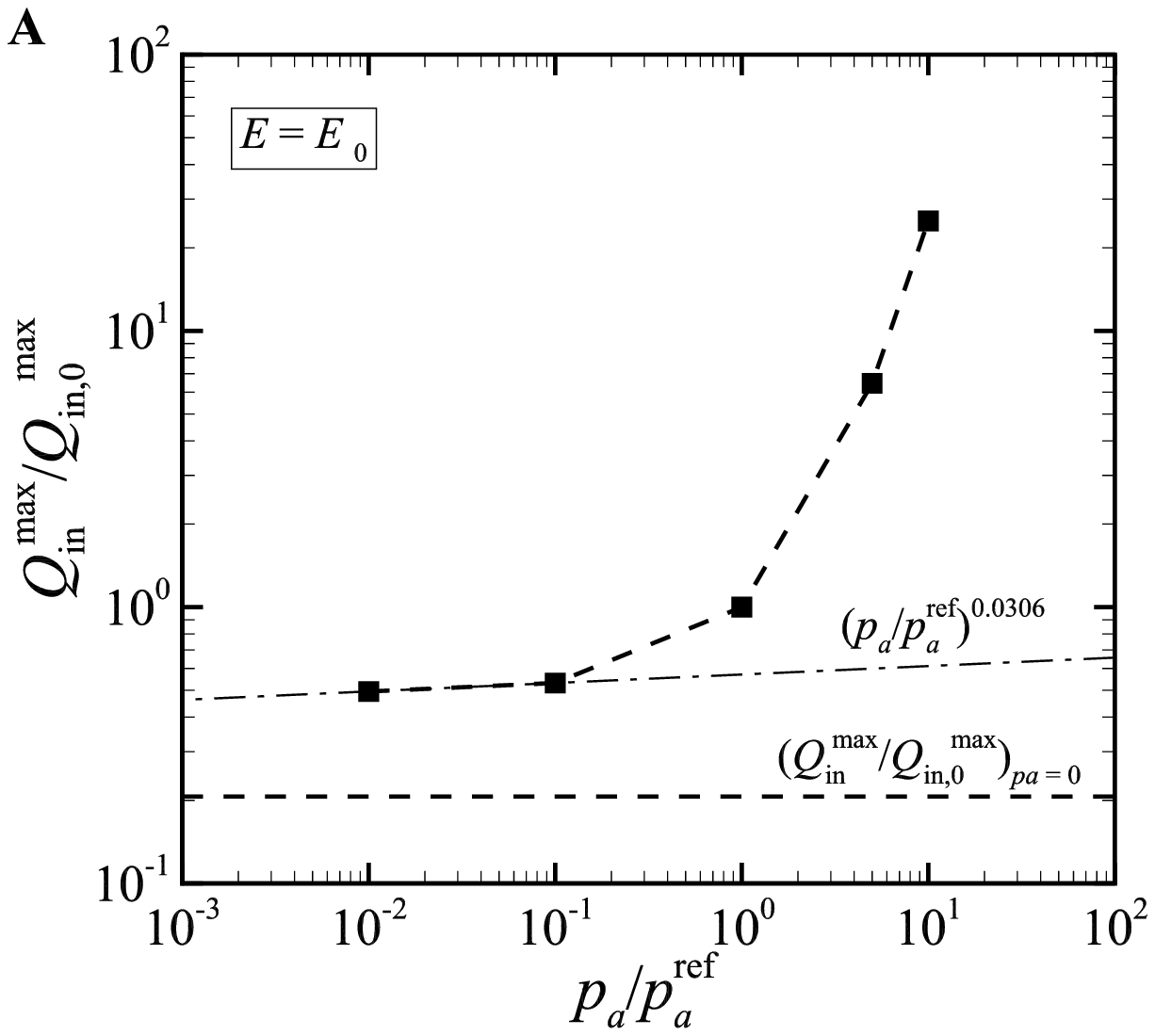}
  \includegraphics[width=8cm]{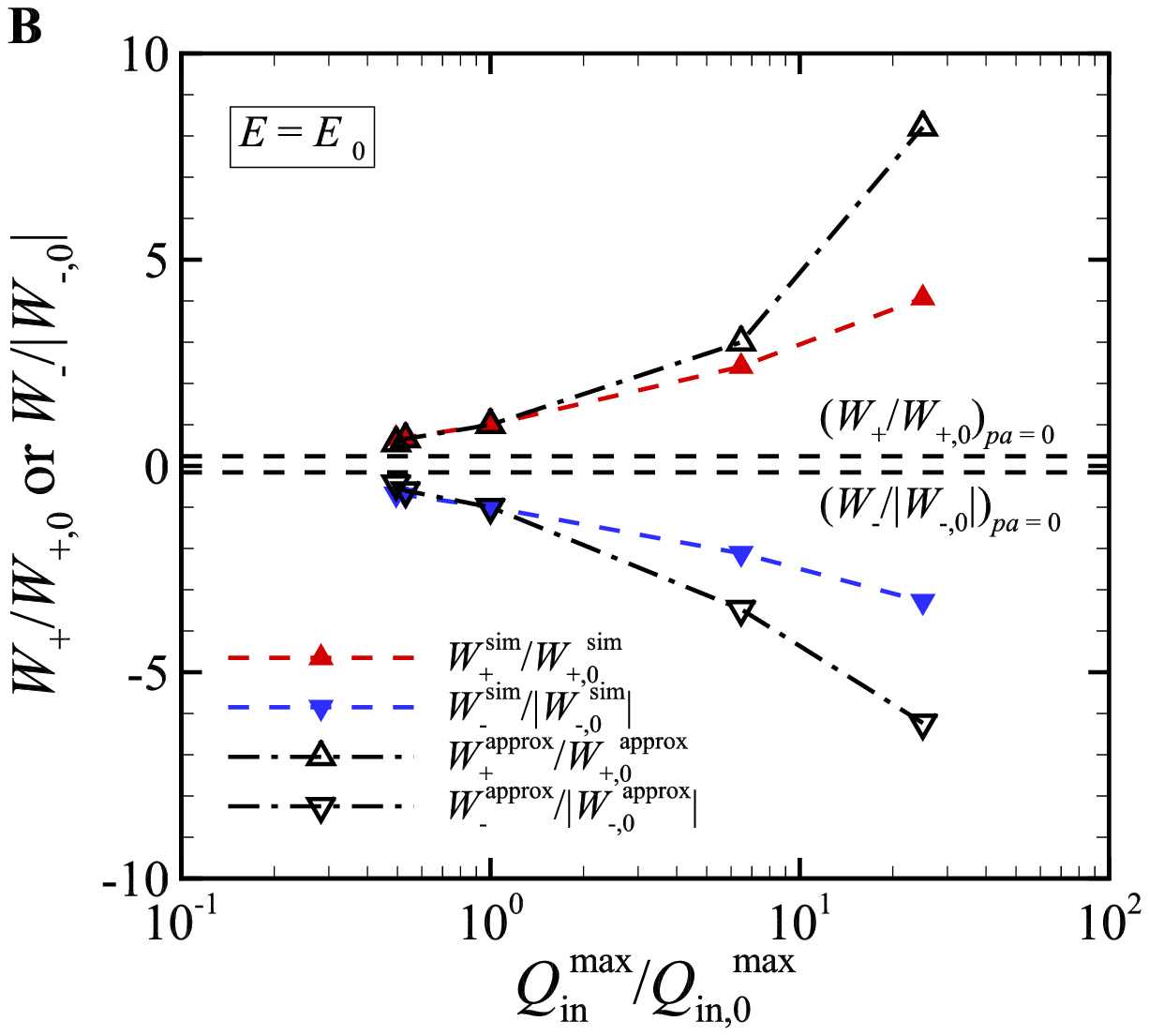}
  \includegraphics[width=8cm]{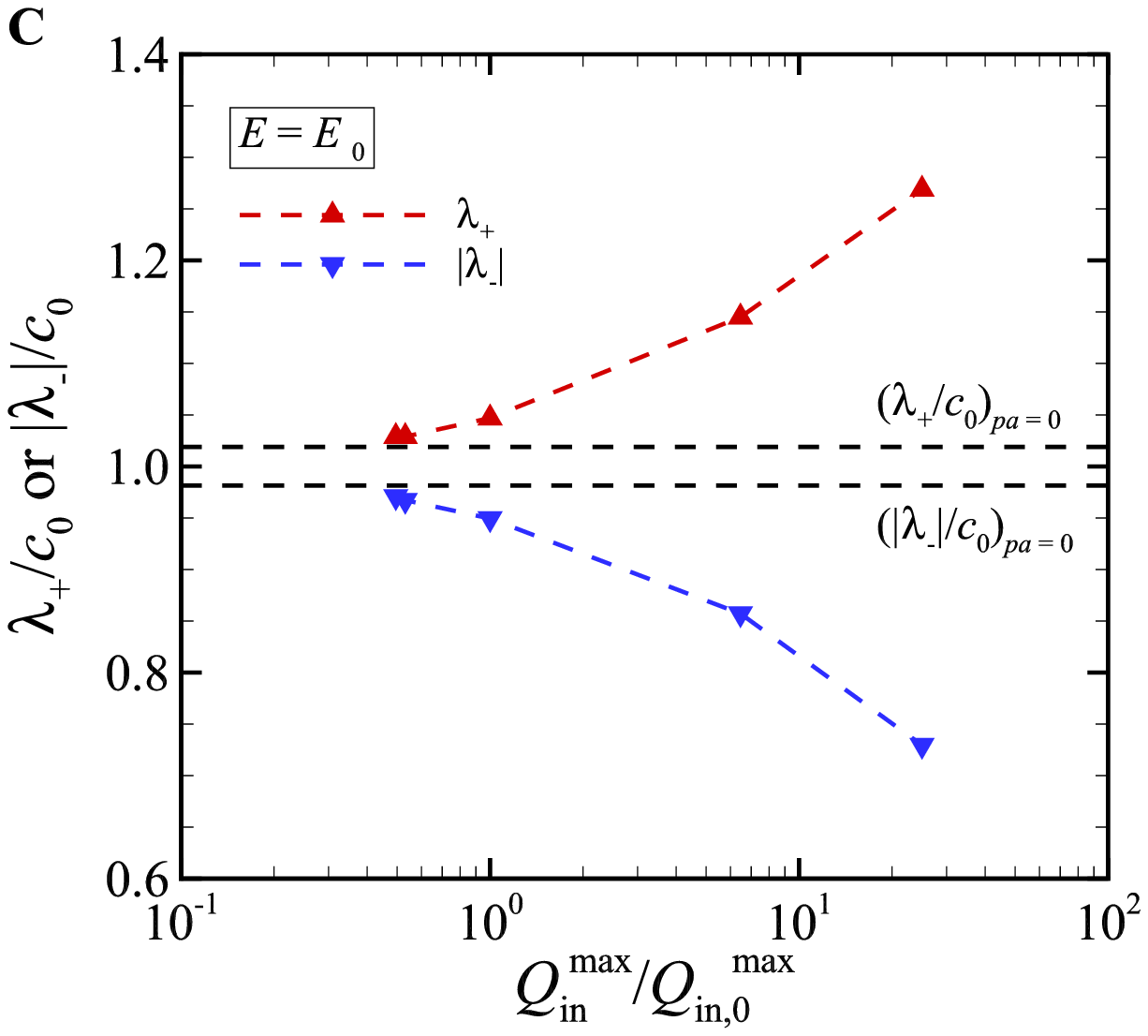}
  \includegraphics[width=8cm]{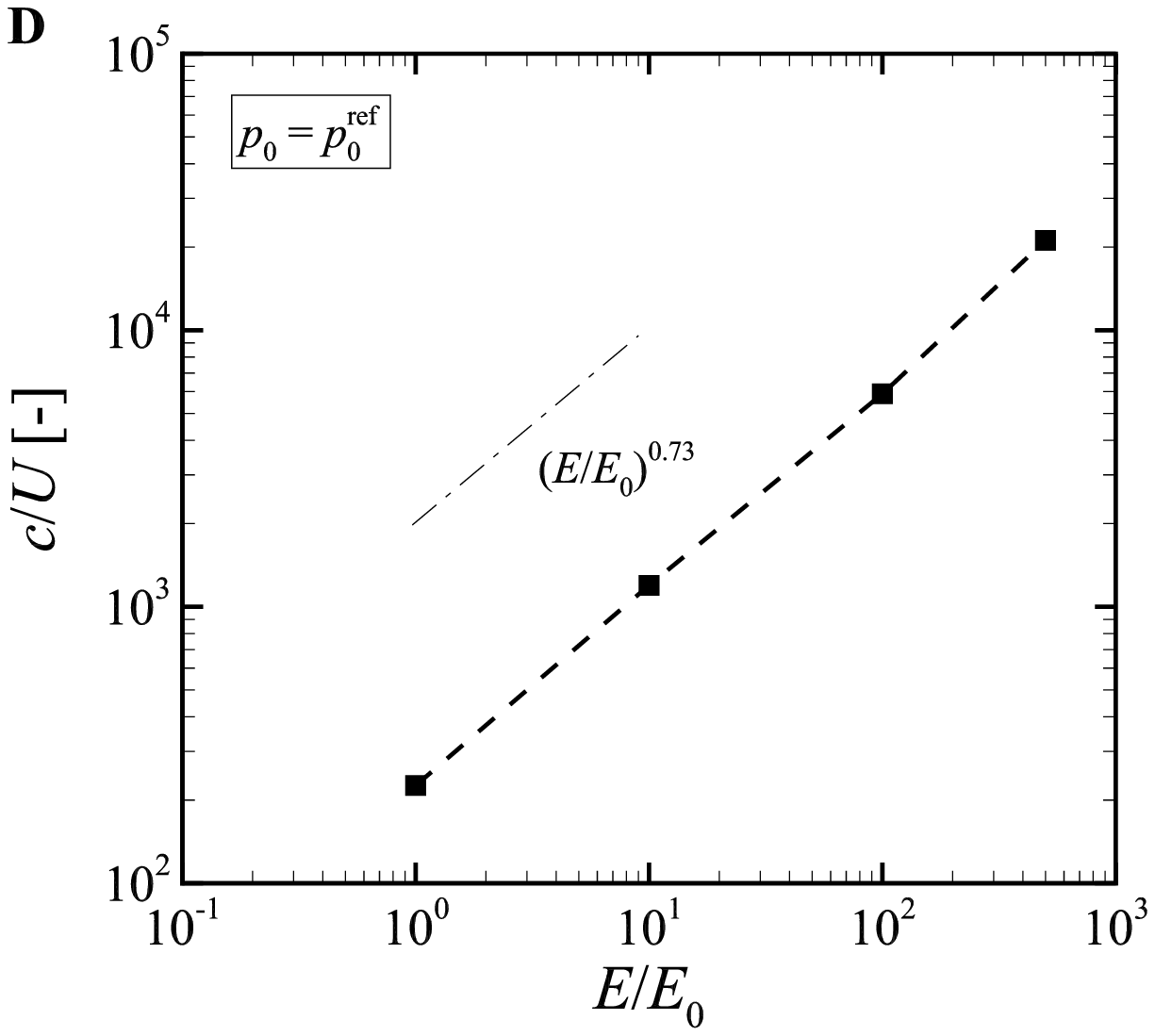}
  \end{center}
  \caption{
  \textbf{(A)} The calculated maximum inlet flow rate ratio $Q_\mathrm{in}^\mathrm{max}/Q_\mathrm{in,0}^\mathrm{max}$ as a function of the maximum inlet pressure $p_a$~\eqref{eq:kinematics},
  where $Q_\mathrm{in,0}^\mathrm{max}$ is the maximum flow rate obtained with the reference pressure amplitude shown in Table~\ref{tab:parameters}, i.e., $p_a^\mathrm{ref} = 4333$ Pa.
  \textbf{(B)} The maximum Riemann invariants $W_\pm$ normalized by those obtained with $p_a^\mathrm{ref}$ and
  \textbf{(C)} the eigen values $\lambda_\pm$ normalized by $c_0$ as a function of the maximum flow rate ratio $Q_\mathrm{in}^\mathrm{max}/Q_\mathrm{in,0}^\mathrm{max}$.
  The results in \textbf{(A)}--\textbf{(C)} were obtained with a Young's modulus of $E_0 = 0.1$ MPa.
  \textbf{(D)} The rate of velocity $U/c$ as a function of the ratio of the Young modulus $E/E_0$ for $p_a^\mathrm{ref}$.
  In \textbf{(A)}--\textbf{(C)}, the results obtained with $p_a = 0$ are also plotted as dashed lines.
  The results in \textbf{(B)}--\textbf{(D)} were quantified at the mid point of a straight tube with length $L = 1$ m and reference radius $r_0 = 10^{-3}$ m.
  The base pressure $p_0$ ($= 10666$ Pa) was same as in the main text.
  }
  \label{fig:verification}
\end{figure*}

To verify the calculated solutions,
the numerical results are compared with those obtained with a weak non-linear equation instead of the theoretical formula.
When $Q = Q_0 + dQ$ and $A = A_0 + dA$,
Equation~\eqref{eq:W_pm} can be written as:
\begin{align}
  W_\pm
  &= \frac{Q_0 + dQ}{A_0 + dA} \pm 4 c_0 \left( \left( 1 + \frac{dA}{A_0} \right)^{1/4} - 1 \right), \\
  &\approx \frac{Q_0}{A_0} \left( 1 - \frac{dA}{A_0} \right) + \frac{dQ}{A_0} \pm c_0 \frac{dA}{A_0}.
\end{align}
When $Q_0 = 0$, we finally have a weak non-linear form of the Riemann invariants~\eqref{eq:W_pm}
\begin{align}
  W_\pm
  &\approx \frac{dQ}{A_0} \pm c_0 \frac{dA}{A_0}.
  \label{eq:weak_W}
\end{align}
Equation~\eqref{eq:weak_W} predicts that wave propagations, including reflective wave $W_-$, are negligible for the small flow rate $dQ \approx 0$ and small vessel contraction $A \approx A_0$.
Indeed, the maximum values of $|W_\pm|$ in the middle point of the straight tube almost collapse on those obtained with $p_a = 0$ when the maximum inlet flow rate $Q_\mathrm{in}^\mathrm{max}$ decreases (Figure~\ref{fig:verification}\textbf{(B)}),
where the tube length and reference radius are set to $L = 1$ m and $r_0 = 10^{-3}$ m, respectively,
and the Young's modulus is $E_0 = 0.1$ MPa,
referring to the value of aortic elasticity in conscious dogs~\cite{Armentano1991}.
The inlet pressure wave form~\eqref{eq:kinematics} and no-reflection condition for the outlet $W_- = 0$ are also considered.
The small maximum inlet inflow rate $Q_\mathrm{in}^\mathrm{max}$ is controlled by the inlet pressure amplitude $p_a$
(see Figure~\ref{fig:verification}\textbf{(A)}). 

Furthermore, the aforementioned weakly non-linear formalization simultaneously leads equation~\eqref{eq:lam_pm} to be approximated as $\lambda_\pm \approx c_{0\pm}$,
and the results are shown in Figure~\ref{fig:verification}\textbf{(C)}.
We also confirmed that the average fluid velocity increased with tube stiffness, as shown in Figure~\ref{fig:verification}\textbf{(D)}.

\printcredits

\bibliographystyle{ieeetr}

\bibliography{cas-refs}

\end{document}